# Switchable half-quantum flux states in a ring of the kagome superconductor $CsV_3Sb_5$

Shuo Wang[1,11], Ilaria Maccari[2,11], Xilin Feng[3], Ze-Nan Wu[4], Jia-Peng Peng[4], Kam Tuen Law[3], Y. X. Zhao[5], Andras Szabo[6], Andreas Schnyder[6], Ning Kang[7], Xiao-Song Wu[8], Jingchao Liu[9], Xuewen Fu[9], Mark H. Fischer[10, †], Manfred Sigrist[2,*], Dapeng Yu[1], Ben-Chuan Lin[1,#]

[1]*International Quantum Academy, and Shenzhen Branch, Hefei National Laboratory, Shenzhen,518048, China.*

[2]*Institute for Theoretical Physics, ETH Zürich, 8093 Zürich, Switzerland.*

[3]*Department of Physics, Hong Kong University of Science and Technology, Clear Water Bay, Hong Kong, China.*

[4]*Southern University of Science and Technology, Shenzhen, 518055, China.*

[5]*Department of Physics and HK Institute of Quantum Science & Technology, The University of Hong Kong, Pokfulam Road, Hong Kong, China.*

[6]*Max-Planck-Institut für Festkörperforschung, Heisenbergstrasse 1, D-70569 Stuttgart, Germany.*

[7]*Key Laboratory for the Physics and Chemistry of Nanodevices, School of Electronics, Peking University, Beijing 100871, China.*

[8]*State Key Laboratory for Artificial Microstructure and Mesoscopic Physics, Frontiers Science Center for Nano-optoelectronics, Peking University, Beijing 100871, China.*

[9]*Ultrafast Electron Microscopy Laboratory, The MOE Key Laboratory of Weak-Light Nonlinear Photonics, School of Physics, Nankai University, Tianjin 300071, China.*

[10]*Department of Physics, University of Zurich, 8057 Zürich, Switzerland.*

[11]*These authors contributed equally to this work.*

[†]mark.fischer@uzh.ch
[*]sigrist@itp.phys.ethz.ch
[#]linbenchuan@iqasz.cn

## Abstract

Magnetic flux quantization in units of $\Phi_0$ = h/2e is a defining feature of superconductivity[1], rooted in the charge-2e nature of Cooper pairs. In a ring geometry, the flux quantization leads to oscillations in the critical temperature with magnetic flux, known as the Little-Parks effect[1–4]. While the maximal critical temperature is conventionally at zero flux, departures from this rule, for instance shifts by a half-quantum flux $\Phi_0/2$, clearly signal unconventional superconducting states[5–10], and requires sign-changing order parameters. Historically, such π-phase shifts in Little-Parks oscillations have been found in tricrystals[6,7] or carefully engineered ring structures[8–10], intentionally incorporating a π-phase shift. Here, we report the discovery of switchable half-quantum flux states in rings made from single crystals of the kagome superconductor[11–15] $CsV_3Sb_5$. We observe Little-Parks oscillations with a π-phase shift at zero bias current, which can be reversibly tuned to conventional Little-Parks oscillations upon applying a bias current. Between the phase change of π- and 0-phase Little-Parks oscillations, h/4e periodic oscillations are observed. Our observations strongly suggest unconventional pairing, potentially in the form of multiple superconducting order parameters, in the kagome superconductor $CsV_3Sb_5$ and reveal a rare, electrically tunable landscape of competing superconducting condensates and fractional flux states.

Superconductors exhibit macroscopic quantum coherence manifesting in phenomena such as persistent currents, flux quantization and coherent phase winding[16]. A central question in modern condensed-matter physics is whether a given superconductor hosts a conventional spin-singlet condensate or an unconventional state[2], for example, with sign-changing pair wave functions. This distinction is fundamental and directly relevant to the prospects for quantum information processing and possible topological quantum computation. Yet most experimental probes primarily access amplitude properties of the superconducting order parameter. A more decisive strategy is to directly interrogate the phase of the condensate[3]. For example, in mesoscopic loop geometries the Little-Parks (LP) effect[1] offers a phase-sensitive probe: the superconducting transition temperature oscillates with magnetic flux due to the quantized phase winding. For a conventional superconductor, the free-energy minima occur at flux $\Phi = n\Phi_0$, producing a resistance minimum at zero field. In contrast, superconductors with sign-changing order parameters[5–10] may acquire an additional $\pi$ phase around the loop, shifting the quantization to $(n+1/2)\Phi_0$ and yielding a $\pi$-phase LP effect. Such half-quantum flux is among the most direct signatures of unconventional pairing symmetry[6–10,17–21], such as d-wave[6–8] or possible p-wave pairing[10,17,18].

The kagome superconductors[11–15] $AV_3Sb_5$ (A = K, Rb, Cs), represent a class of layered materials with intriguing electronic topology and correlated phenomena. These compounds are characterized by alternating stacks of vanadium-antimony (V-Sb) kagome layers, interleaved with planar Sb layers and alkali-metal (A) layers. Among this class, $CsV_3Sb_5$ has drawn particular attention due to its relatively high crystallinity and the highest superconducting critical temperature of $T_c \sim 2.5$ K[13,15]. Structurally, each V-Sb layer features a two-dimensional (2D) kagome lattice of vanadium atoms, forming a hexagonal structure with Sb atoms occupying the hexagon centers, as schematically illustrated in Fig. 1a. The kagome structure of this superconductor naturally hosts correlated phenomena due to its unique band structure and the Fermi level near a Van Hove singularity. Indeed, the system undergoes a charge-order transition with a 2x2 in-plane reconstruction at approximately 94 K[22–33] , which is suggested to result in a host of complex phenomena, many still under active debate, from chiral order[22,26,33,34], nematicity[27,28,31,32], to time-reversal-symmetry breaking[22,26,28,35–38], and many-body coherence[39]. While the superconducting state emerging out of this exotic phase shows signs of being conventional – Knight shift[40] suggests possible spin-singlet pairing and several groups have reported a full pairing gap[38,41–43] – other experiments hint at anisotropic two-gap superconducvity[44], two-dome superconductivity under pressure[29,45], time-reversal-symmetry breaking[38,46–48] superconductivity, the existence of a pair-density wave[39–42], and even higher-charge condensates[49]. Despite the host of experiments, the nature of the superconducting state in $CsV_3Sb_5$ thus remains controversial, including whether the pairing is unconventional[13,50,51].

Here, we report the realization of switchable half-quantum flux states in the kagome superconductor $CsV_3Sb_5$. By engineering mesoscopic ring structures from thin $CsV_3Sb_5$ flakes, a π-phase Little-Parks effect is observed. Moreover, a current-driven transition from π-phase to the zero-phase LP state emerges. Strikingly, between the two phases, a regime of $\Phi_0/2$ periodicity, namely h/4e periodicity, is observed. Our findings provide compelling evidence for unconventional pairing, potentially rooted in multiple, (nearly) degenerate superconducting orders. The demonstrated all-electric control establishes $CsV_3Sb_5$ as a unique experimental platform for probing the interplay between unconventional superconducting pairing, fractional quantum fluxes, and possible higher-order condensates.

**Device Configuration**

To investigate the superconducting ground state and flux-quantization behavior of $CsV_3Sb_5$, we fabricated mesoscopic rings from exfoliated thin flakes. A total of 13 devices were fabricated and systematically measured, with full details provided in the Supplementary Information. The optical image of a typical device is shown in Fig. 1b, with an effective inner ring area of approximately 1 $\mu m^2$ with the arm width of 200 nm. The resistance-temperature (R-T) curve exhibits a typical two-step drop as the system transitions into the superconducting state, as shown in Fig. 1b. The initial drop corresponds to the superconducting transition of the unpatterned flake, while the second drop is attributed to the ring structure, where geometric confinement suppresses the superconducting onset temperature. In Fig. 1b, the blue region refers to the regime near the superconducting critical temperature, where the main finding, namely the half-quantum flux, was found. Fig. 1c shows the differential resistance versus bias current at different temperatures. With temperature increasing, the zero-resistance regime shrinks, showing a typical superconducting behavior. The evolution of out-of-plane magneto-resistance of the mesoscopic ring on temperatures is presented in Fig. 1d, consistent with the superconducting transition in Fig. 1b. The Supplementary Information presents the details of pristine $CsV_3Sb_5$ thin flakes and device fabrication methods.

## Little-Parks effect with a π phase shift

The Little-Parks effect provides an insightful experiment to probe unconventional superconducting states, possibly revealing their non-trivial phase structure. In a superconducting annulus, the total magnetic flux threading the loop—known as the fluxoid Φ—is quantized according to the relation[9]

$$\Phi = \Phi_{ext} + 4\pi \oint \lambda^2 \vec{j_s} \cdot \vec{ds} = n\,\Phi_0,$$

where $\vec{j}_s$ is the supercurrent density, $\Phi_{ext}$ is the applied external magnetic flux through the ring, and λ is the London penetration depth. This quantization condition ensures that the superconducting wavefunction remains single-valued around the loop.

As the external magnetic field is varied, the supercurrent adjusts to preserve this quantization, leading to an increase in the free energy and thus a suppression of $T_c$. However, once a flux quantum is threaded through the ring, no current is required and, as a result, a periodic modulation of the superconducting free energy occurs. This can be captured by a free-energy expansion of the loop, $\Delta F(\Phi) = E_1\cos(2\pi\Phi/\Phi_0)$ with $E_1<0$. Consequently, the superconducting transition temperature $T_c$ – defined as the temperature at which the free energies of the normal and superconducting states become equal – exhibits a periodic dependence on the enclosed magnetic flux. This phenomenon underpins the Little-Parks effect, in which $T_c$ oscillates with a period corresponding to the superconducting flux quantum $\Phi_0 = h/2e$. At temperatures near $T_c$, where the system is near the onset of superconductivity, these periodic changes in free energy can be sensitively probed via magnetoresistance measurements. In this regime, resistance variations serve as an effective proxy for $T_c$ oscillations, enabling the extracting of phase-sensitive information related to flux quantization. This resistance-based approach[1] thus provides a powerful experimental window into the underlying mechanism of the superconducting order parameter.

In conventional superconductors, resistance minima at $\Phi = n\Phi_0$ appear near the critical temperature, reflecting corresponding maxima in $T_c$. In contrast, unconventional superconductors, whose pair wave functions typically exhibit non-trivial angular phase dependence, can be structured in a way, for example using Josephson junctions[6–8,18], to incorporate a π-phase shift within a superconducting loop. This shift produces the resistance minima at $(n+1/2)\Phi_0$, as illustrated in Fig. 2a.

Such a shift can also occur in polycrystalline samples, where sign-changing order parameters can accidentally lead to frustration in the form of a π-phase shift[10]. Finally, in samples derived from the transition metal dichalcogenide superconductor $2H\text{-}TaS_2$[20], π-phase shifts were found even for single crystals. Importantly, in both polycrystalline samples where grain orientations and junction phases are distributed randomly, and the $2H\text{-}TaS_2$ derivatives, only a fraction of the rings shows a π-phase shift, underscoring the statistical rather than systematic nature of the origin of the shift. Note that we can again capture a π-phase shift in the periodic modulation of the superconducting free energy by setting $E_1 > 0$, which shifts the minima from integer to half-integer flux quanta.

The resistance oscillations of the thin-flake $CsV_3Sb_5$ mesoscopic rings with constant periods near the critical temperature $T_c$ is presented in Fig. 2b. The variation of $T_c$ manifests as resistance oscillations in the Little-Parks effect. After subtracting a polynomial even-function fit ($R= R_2*B^2+R_0$) of the background signal, the oscillating part ΔR is shown as a waterfall plot in Fig. 2c. Details of the background-signal extraction are shown in Fig. S10. The corresponding color map of the oscillating magnetoresistance ΔR at different temperatures is shown in Fig. 2d. In stark contrast to conventional superconductors, we observe a resistance peak at zero magnetic field. Specifically, we find a clear periodic behavior with resistance minimum at $\Phi = (n+1/2)\Phi_0$ rather than $n\Phi_0$, consistent with the π-phase Little-Parks oscillations. The oscillation period is around 18 Oe, consistent with the ring geometry. The π-phase shift persists from around 2.9 K to the highest temperature of 3.25 K, which is near the critical temperature, corresponding to the blue region in Fig. 1b. With the temperature further increasing, the oscillations disappear due to the loss of coherence over the length scale of the ring device. Importantly, we emphasize that almost all of the measured 13 samples, namely 12, intrinsically show the π-phase Little-Parks effect. The statistics and the measurement details of the Little-Parks oscillations can be found in the Supplementary Information (See Supplementary Note 1 and Figs. S1~S8).

## Bias current-driven phase transition

Motivated by the idea that a finite supercurrent can couple to the internal phase structure of competing superconducting configurations, we next study how the Little–Parks oscillations evolve under a direct current (DC) bias. The AC is maintained at a constant 1 μA, while the DC is varied continuously. Fig. 3a shows the mapping plot of the bias-current-driven phase transition of the unconventional π-phase Little-Parks effect (π-LP) to the conventional Little-Parks (0-LP) effect at a temperature of 2.8 K. The extracted periodic oscillations at different bias currents are presented in Fig. 3b. Notably, the original nontrivial resistance minimum at $\Phi = (n+1/2)\Phi_0$ transits to the trivial resistance minimum at $\Phi = n\Phi_0$ at high bias currents, here around 18 μA. Fig. 3c further illustrates this with a waterfall plot of the extracted resistance oscillations.

A noteworthy feature arises in the vicinity of the transition from π-phase to 0-phase Little-Parks (LP) oscillations: the emergence of half-period (h/4e) oscillations. A detailed analysis of this intermediate regime is shown in Fig. 4, which captures the evolution of the LP oscillation period under varying bias current, ranging from 16 μA to 18 μA. As shown, the system evolves from integer-period (h/2e) LP oscillations with a π-phase shift, to half-integer period (h/4e) oscillations, and finally returns to integer-period (h/2e) LP oscillations without the π-phase shift. Fig. 4a presents the extracted magnetoresistance mapping as a function of bias current. The dashed line corresponds to h/4e period. The corresponding waterfall plot is shown in Fig. 4b. The half-integer oscillations with a period of h/4e become most pronounced at a bias current of 17.2 μA. To quantitatively assess this evolution, we performed a fast Fourier transform (FFT) analysis, as shown in Fig. 4c. The relative amplitude of the h/4e component compared to the

h/2e component, expressed as $A_{FFT}(4e)/A_{FFT}(2e)$ or $A_{\Phi_0/2}/A_{\Phi_0}$, serves as a measure of the dominant oscillation mode. This ratio reaches a maximum near 17.2 μA, consistent with the features observed in the waterfall plot in Fig. 4b and reinforcing the presence of the intermediate half-period regime. These findings provide compelling evidence for the transient formation of h/4e periods during the phase evolution, manifesting as a distinct h/4e interference pattern in the LP oscillations. We also observed a similar phase evolution for another temperature T = 2.5 K with the critical value of the phase evolution changed, as shown in Fig. S12. Fig. S13 further shows the phase diagram in another device #5.

## Discussion

The π-phase Little-Parks effect provides compelling evidence for unconventional superconductivity in $CsV_3Sb_5$ characterized by a sign-changing order parameter[6–10,17–20]. Central to this phenomenon is the acquisition of a π phase upon going around the ring, which is fundamentally incompatible with the phase structure of conventional s-wave superconductors. Such a π-phase shift requires a sign change in the superconducting order parameter. In d-wave superconductors, for instance, the order parameter changes sign upon a 90° rotation, while in p-wave superconductors, it reverses under a 180° rotation. Consequently, π-phase shifts have been observed in d-wave superconducting tricrystals[6,7], in superconductors employing carefully engineered ring structures[9,10,17–20], or rings made from polycrystalline samples. Moreover, even in ring structures made from single-crystal samples, π-phase shifts have been recently reported in superconductors derived from 2H-$TaS_2$[19,20]. Importantly, for all systems except composite rings, only a fraction of rings show a π-phase shift, while the others show the conventional behavior[10,17,20]. As such, any scenario discussed for these systems is not fully compatible with our observation of π-phase shifts in almost all of our samples (12 out of 13).

In order to further discuss possible scenarios, we reiterate our main findings, namely (1) a preponderance of rings made from the kagome superconductor $CsV_3Sb_5$ showing a π-phase shift, (2) a switch from π phase to 0 phase upon increasing the bias current, and finally (3), the observation of half-quantum-flux periodicity at intermediate bias currents. Any proposal for a superconducting order parameter in $CsV_3Sb_5$ has to address these points.

While the ring-structure samples are made of single crystals, the fabrication process could still induce lattice defects or effective polycrystallinity, leading to an overall frustration of the phase going around the ring. However, as already discussed, in this case we would expect only a fraction of the rings to host a π-phase shift, which is inconsistent with our findings of nearly all devices displaying π-phase behavior. Moreover, this explanation does not naturally account for the observed current-driven evolution from a π-phase to a 0-phase state.

In the presence of sample specifics such as strain fields, geometric confinement, or special defects[21], a multi-component superconducting order parameter could favor a π-phase shift when going around the ring, despite the energetic cost associated with the phase winding[19,20]. In such a scenario, a subdominant order parameter without phase winding might be in close energetic proximity. Yet, also in this second scenario, we would not expect a preponderance of π-phase shifts.

These considerations then point toward either an intrinsic mechanism or a systematic aspect of the fabrication process. Despite the low level of disorder in the sample, as reflected in the sharp resistive transition, systematic aspects like a mismatch between the kagome lattice with the substrate could indeed induce non-trivial strain fields, which may couple with a two-component superconducting order parameter and thereby lead to the observed π-phase shift[20]. In addition, the nematic order[27,28,31,32], suggested to exist above the superconducting critical temperature, could enhance any effect of strain, to which $CsV_3Sb_5$ is known to be extremely susceptible[32].

The second key finding of our work is the systematic switch from a π-phase to a 0-phase ring upon increasing the bias current. Such a π-switch has previously been reported only as a function of the applied magnetic field[52]. The theoretical interpretation invoked a dipole-type spin-orbit interaction, which, under a finite bias current, can generate a non-trivial spin texture in spin-triplet unconventional superconductors and thus a π-phase shift in the oscillations[52–54]. Alternatively, we suggest that in a spin-singlet multicomponent superconductor, a finite bias current can affect the components differently, possibly driving a subdominant order parameter to become dominant above a current threshold. If the dominant superconducting order parameter acquires a π-phase going around the ring while the subdominant does not, the applied current can induce a switch between a π-phase and a 0-phase ring. We elaborate on this scenario in the Supplementary Information.

We now address the third central finding of our work, the observation of h/4e oscillations, clearly visible at low magnetic fields and at intermediate bias currents. Such oscillations are usually interpreted as arising from a phase coherent state formed by four electrons rather than two, leading to a charge-4e condensate. These higher-order electron condensates may occur in multiband superconductors[55,56] or in systems with multicomponent order parameters, and are stabilized by topological excitations of the phase at finite temperature[57,58]. As such, charge-4e superconducting states can arise from the partial melting of pair-density-wave (PDW)[59–62] or nematic ground states[63–65], and have even been discussed recently in the context of kagome-lattice systems[66,67]. In our case, however, we believe the h/4e-like oscillations in Fig. 4 have a different origin: With the oscillations emerging only between the π-phase LP oscillations and 0-phase LP oscillations at finite bias currents, a natural explanation is provided by a current-induced switching between a dominant superconducting order parameter that changes sign along the ring and a subdominant uniform order parameter. In this picture, the h/4e signal arises from superimposing π-phase and 0-phase Little-Parks oscillations, whose relative weight evolves gradually with bias current. The continuous crossover observed in Fig. 4 supports this interpretation. This picture can be rationalized within the free-energy expansion introduced above,

$$\Delta F(\Phi) = E_1 \cos(2\pi\Phi/\Phi_0) + E_2 \cos(4\pi\Phi/\Phi_0) + \cdots,$$

where $E_1$ evolves from being positive at low current, thus exhibiting π-shifted oscillations, to being negative at larger currents. In addition to the first harmonic term ($E_1$), we have introduced a second harmonic ($E_2$), generically allowed by symmetry. While the former term is usually dominant, the latter one takes over when $E_1$ continuously changes sign, giving rise to the h/4e periodicity. Thus, we attribute the observed h/4e-like oscillations to higher-order harmonic contributions in the free energy. For a motivation of such a periodicity of the free energy from a Ginzburg-Landau-theory perspective, we refer to the Supplementary Information. While we believe the above scenario for the h/4e oscillations is the most natural, we cannot exclude other scenarios including genuine charge-4e superconducting states stabilized by a finite bias current, or more exotic origins of the π-phase shift, like the unconventional normal state of $CsV_3Sb_5$[13,22,34,39].

In summary, we report a consecutive phase modulation of half-integer quantum flux, h/4e-period interference pattern and integer quantum flux in the mesoscopic rings of the kagome superconductor $CsV_3Sb_5$. The observations are consistent with sign-changing superconducting order parameters. Regardless of the specific physical mechanism, such electric controllability of 0-phase and π-phase superconducting devices without any heterojunction design[68,69] holds promise for ultralow-power, next-generation integrated circuits, such as cryogenic memory, Josephson phase battery[70] or superconducting programmable logic circuits[71–73]. Such all-electrically controlled transitions also provide a way to electrically manipulate and engineer these unconventional superconducting states, which may be helpful to possible applications such as topological qubits[74–76] and flux qubits[77,78].

# Methods

## Sample fabrication

The single crystal $CsV_3Sb_5$ was grown via the conventional flux method. The thin flakes of $CsV_3Sb_5$ were exfoliated using polydimethylsiloxane (PDMS) from the bulk crystal onto silicon substrates with a 285 nm oxide layer. The fabrication of metal electrodes was achieved using the standard electron-beam lithography (EBL) method, followed by deposition of a Pd/Au (10 nm/50 nm) through sputtering evaporation. Then, the device was encapsulated with hBN to prevent any degradation due to air atmosphere. A full description of the etching procedures and the Little-Parks measurement methodology is available in the Supplementary Information. All device pre-fabrication procedures mentioned above were performed in a controlled environment within a nitrogen-filled glovebox with oxygen and water levels below 0.01 ppm to minimize any potential degradation of the samples.

## Transport measurements

The samples were measured in an Oxford dilution refrigerator Triton XL1000. To optimize the signal-to-noise ratio in these experiments, a standard lock-in technique was employed, along with the implementation of essential low-temperature filters.

# Data availability

The data that support the plots within this paper and other related findings are available from the corresponding author upon request. Alternatively, the data are accessible via the Figshare repository.

# References

1. Little, W. A. & Parks, R. D. Observation of Quantum Periodicity in the Transition Temperature of a Superconducting Cylinder. *Phys. Rev. Lett.* **9**, 9–12 (1962).
2. Sigrist, M. & Ueda, K. Phenomenological theory of unconventional superconductivity. *Rev. Mod. Phys.* **63**, 239–311 (1991).
3. Van Harlingen, D. J. Phase-sensitive tests of the symmetry of the pairing state in the high-temperature superconductors—Evidence for $d_{x^2-y^2}$ symmetry. *Rev. Mod. Phys.* **67**, 515–535 (1995).
4. Tsuei, C. C. & Kirtley, J. R. Pairing symmetry in cuprate superconductors. *Rev. Mod. Phys.* **72**, 969–1016 (2000).
5. Leggett, A. J. A theoretical description of the new phases of liquid $^3$He. *Rev. Mod. Phys.* **47**, 331–414 (1975).
6. Geshkenbein, V. B., Larkin, A. I. & Barone, A. Vortices with half magnetic flux quanta in '"heavy-fermion"' superconductors. *Phys. Rev. B* **36**, 235–238 (1987).
7. Tsuei, C. C. *et al.* Pairing Symmetry and Flux Quantization in a Tricrystal Superconducting Ring of $YBa_2Cu_3O_{7-\delta}$. *Phys. Rev. Lett.* **73**, 593–596 (1994).
8. Wollman, D. A., Van Harlingen, D. J., Lee, W. C., Ginsberg, D. M. & Leggett, A. J. Experimental determination of the superconducting pairing state in YBCO from the phase coherence of YBCO-Pb dc SQUIDs. *Phys. Rev. Lett.* **71**, 2134–2137 (1993).
9. Jang, J. *et al.* Observation of Half-Height Magnetization Steps in $Sr_2RuO_4$. *Science* **331**, 186–188 (2011).
10. Li, Y., Xu, X., Lee, M.-H., Chu, M.-W. & Chien, C. L. Observation of half-quantum flux in the unconventional superconductor β-$Bi_2Pd$. *Science* **366**, 238–241 (2019).
11. Ortiz, B. R. *et al.* New kagome prototype materials: discovery of $KV_3Sb_5$, $RbV_3Sb_5$, and $CsV_3Sb_5$. *Phys. Rev. Materials* **3**, 094407 (2019).
12. Ortiz, B. R. *et al.* $CsV_3Sb_5$ : A $Z_2$ Topological Kagome Metal with a Superconducting Ground State. *Phys. Rev. Lett.* **125**, 247002 (2020).
13. Wang, Y. Quantum states and intertwining phases in kagome materials. *nature reviews physics* **5**, 635–658 (2023).
14. Jiang, K. *et al.* Kagome superconductors $AV_3Sb_5$(A = K, Rb, Cs). *National Science Review* **10**, nwac199 (2023).
15. Wilson, S. D. & Ortiz, B. R. $AV_3Sb_5$ kagome superconductors. *Nat Rev Mater* **9**, 420–432 (2024).
16. Michael Tinkham. *Introduction to Superconductivity*.
17. Xu, X., Li, Y. & Chien, C. L. Spin-Triplet Pairing State Evidenced by Half-Quantum Flux in a Noncentrosymmetric Superconductor. *Phys. Rev. Lett.* **124**, 167001 (2020).

18. Xu, X., Li, Y. & Chien, C. L. Observation of Odd-Parity Superconductivity with the Geshkenbein-Larkin-Barone Composite Rings. *Phys. Rev. Lett.* **132**, 056001 (2024).

19. Wan, Z. *et al.* Unconventional superconductivity in chiral molecule–$TaS_2$ hybrid superlattices. *Nature* **632**, 69–74 (2024).

20. Almoalem, A. *et al.* The observation of π-shifts in the Little-Parks effect in 4Hb-$TaS_2$. *Nat Commun* **15**, 4623 (2024).

21. Fischer, M. H., Lee, P. A. & Ruhman, J. Mechanism for π phase shifts in Little-Parks experiments: Application to 4Hb−$TaS_2$ and to 2H−$TaS_2$ intercalated with chiral molecules. *Phys. Rev. B* **108**, L180505 (2023).

22. Jiang, Y.-X. *et al.* Unconventional chiral charge order in kagome superconductor $KV_3Sb_5$. *Nat. Mater.* **20**, 1353–1357 (2021).

23. Zhao, H. *et al.* Cascade of correlated electron states in the kagome superconductor $CsV_3Sb_5$. *Nature* **599**, 216–221 (2021).

24. Kang, M. *et al.* Charge order landscape and competition with superconductivity in kagome metals. *Nat. Mater.* **22**, 186–193 (2022).

25. Hu, Y. *et al.* Coexistence of trihexagonal and star-of-David pattern in the charge density wave of the kagome superconductor $AV_3Sb_5$. *Phys. Rev. B* **106**, (2022).

26. Li, H. *et al.* Rotation symmetry breaking in the normal state of a kagome superconductor $KV_3Sb_5$. *Nat. Phys.* **18**, 265–270 (2022).

27. Nie, L. *et al.* Charge-density-wave-driven electronic nematicity in a kagome superconductor. *Nature* **604**, 59–64 (2022).

28. Xu, Y. *et al.* Three-state nematicity and magneto-optical Kerr effect in the charge density waves in kagome superconductors. *Nat. Phys.* **18**, 1470–1475 (2022).

29. Zheng, L. *et al.* Emergent charge order in pressurized kagome superconductor $CsV_3Sb_5$. *Nature* **611**, 682–687 (2022).

30. Saykin, D. R. *et al.* High Resolution Polar Kerr Effect Studies of $CsV_3Sb_5$: Tests for Time-Reversal Symmetry Breaking below the Charge-Order Transition. *Phys. Rev. Lett.* **131**, 016901 (2023).

31. Asaba, T. *et al.* Evidence for an odd-parity nematic phase above the charge-density-wave transition in a kagome metal. *Nat. Phys.* **20**, 40–46 (2024).

32. Guo, C. *et al.* Correlated order at the tipping point in the kagome metal $CsV_3Sb_5$. *Nat. Phys.* **20**, 579–584 (2024).

33. Xing, Y. *et al.* Optical manipulation of the charge-density-wave state in $RbV_3Sb_5$. *Nature* **631**, 60–66 (2024).

34. Guo, C. *et al.* Switchable chiral transport in charge-ordered kagome metal $CsV_3Sb_5$. *Nature* **611**, 461–466 (2022).

35. Feng, X., Jiang, K., Wang, Z. & Hu, J. Chiral flux phase in the Kagome superconductor $AV_3Sb_5$. *Science Bulletin* **66**, 1384–1388 (2021).

36. Khasanov, R. *et al.* Time-reversal symmetry broken by charge order in $CsV_3Sb_5$. *Phys. Rev. Research* **4**, 023244 (2022).

37. Mielke, C. *et al.* Time-reversal symmetry-breaking charge order in a kagome superconductor. *Nature* **602**, 245–250 (2022).

38. Guguchia, Z. *et al.* Tunable unconventional kagome superconductivity in charge ordered $RbV_3Sb_5$ and $KV_3Sb_5$. *Nat Commun* **14**, 153 (2023).

39. Guo, C. *et al.* Many-body interference in kagome crystals. *Nature* **647**, 68–73 (2025).

40. Mu, C. *et al.* S-Wave Superconductivity in Kagome Metal $CsV_3Sb_5$ Revealed by $^{121/123}$Sb NQR and $^{51}$V NMR Measurements. *Chinese Phys. Lett.* **38**, 077402 (2021).

41. Duan, W. *et al.* Nodeless superconductivity in the kagome metal $CsV_3Sb_5$. *Sci. China Phys. Mech. Astron.* **64**, 107462 (2021).

42. Gupta, R. *et al.* Microscopic evidence for anisotropic multigap superconductivity in the $CsV_3Sb_5$ kagome superconductor. *npj Quantum Mater.* **7**, 49 (2022).

43. Zhong, Y. *et al.* Nodeless electron pairing in $CsV_3Sb_5$-derived kagome superconductors. *Nature* **617**, 488–492 (2023).

44. Hossain, M. S. *et al.* Unconventional gapping behaviour in a kagome superconductor. *Nat. Phys.* **21**, 556–563 (2025).

45. Chen, K. Y. *et al.* Double Superconducting Dome and Triple Enhancement of Tc in the Kagome Superconductor $CsV_3Sb_5$ under High Pressure. *Phys. Rev. Lett.* **126**, 247001 (2021).

46. Wang, S. *et al.* Spin-polarized p-wave superconductivity in the kagome material $RbV_3Sb_5$. Preprint at http://arxiv.org/abs/2405.12592 (2024).

47. Le, T. *et al.* Superconducting diode effect and interference patterns in kagome $CsV_3Sb_5$. *Nature* **630**, 64–69 (2024).

48. Deng, H. *et al.* Evidence for time-reversal symmetry-breaking kagome superconductivity. *Nat. Mater.* **23**, 1639–1644 (2024).

49. Ge, J. *et al.* Charge-4e and Charge-6e Flux Quantization and Higher Charge Superconductivity in Kagome Superconductor Ring Devices. *Phys. Rev. X* **14**, 021025 (2024).

50. Wu, X. *et al.* Nature of Unconventional Pairing in the Kagome Superconductors $AV_3Sb_5$ (A = K, Rb, Cs). *Phys. Rev. Lett.* **127**, 177001 (2021).

51. Dai, Y. Existence of Hebel-Slichter peak in unconventional kagome superconductors. *Physical Review B* **110**, 144516 (2024).

52. Niimi, Y. *et al.* Observation of the crossover from quantum fluxoid to half-quantum fluxoid in a chiral superconducting device. *Science Advances* **11**, eadw6625 (2025).

53. Aoyama, K. Little-Parks oscillation and d -vector texture in spin-triplet superconducting rings with bias current. *Phys. Rev. B* **106**, L060502 (2022).

54. Aoyama, K. Half-quantum flux in spin-triplet superconducting rings with bias current. *Phys. Rev. B* **108**, L060502 (2023).

55. Grinenko, V. *et al.* State with spontaneously broken time-reversal symmetry above the superconducting phase transition. *Nat. Phys.* **17**, 1254–1259 (2021).

56. Shipulin, I. *et al.* Calorimetric evidence for two phase transitions in $Ba_{1-x}K_xFe_2As_2$ with fermion pairing and quadrupling states. *Nat Commun* **14**, 6734 (2023).

57. Babaev, E. Phase diagram of planar U(1)×U(1) superconductor. *Nuclear Physics B* **686**, 397–412 (2004).

58. Babaev, E., Sudbø, A. & Ashcroft, N. W. A superconductor to superfluid phase transition in liquid metallic hydrogen. *Nature* **431**, 666–668 (2004).

59. Agterberg, D. F. & Tsunetsugu, H. Dislocations and vortices in pair-density-wave superconductors. *Nature Phys* **4**, 639–642 (2008).

60. Berg, E., Fradkin, E. & Kivelson, S. A. Charge-4e superconductivity from pair-density-wave order in certain high-temperature superconductors. *Nature Phys* **5**, 830–833 (2009).

61. Agterberg, D. F., Geracie, M. & Tsunetsugu, H. Conventional and charge-six superfluids from melting hexagonal Fulde-Ferrell-Larkin-Ovchinnikov phases in two dimensions. *Phys. Rev. B* **84**, 014513 (2011).

62. Agterberg, D. F. *et al.* The Physics of Pair-Density Waves: Cuprate Superconductors and Beyond. *Annu. Rev. Condens. Matter Phys.* **11**, 231–270 (2020).

63. Fernandes, R. M. & Fu, L. Charge-4 e Superconductivity from Multicomponent Nematic Pairing: Application to Twisted Bilayer Graphene. *Phys. Rev. Lett.* **127**, 047001 (2021).

64. Jian, S.-K., Huang, Y. & Yao, H. Charge-4e Superconductivity from Nematic Superconductors in Two and Three Dimensions. *Phys. Rev. Lett.* **127**, 227001 (2021).

65. Zou, X., Wan, Z.-Q. & Yao, H. Emergence of charge-4e superconductivity from 2D nematic superconductors. Preprint at https://doi.org/10.48550/arXiv.2510.26720 (2025).

66. Zhou, S. & Wang, Z. Chern Fermi pocket, topological pair density wave, and charge-4e and charge-6e superconductivity in kagomé superconductors. *Nat Commun* **13**, 7288 (2022).

67. Varma, C. M. & Wang, Z. Extended superconducting fluctuation region and 6 e and 4 e flux quantization in a kagome compound with a normal state of 3 Q order. *Phys. Rev. B* **108**, 214516 (2023).

68. Linder, J. & Robinson, J. W. A. Superconducting spintronics. *Nature Phys* **11**, 307–315 (2015).

69. Gingrich, E. C. *et al.* Controllable 0–π Josephson junctions containing a ferromagnetic spin valve. *Nature Phys* **12**, 564–567 (2016).

70. Strambini, E. *et al.* A Josephson phase battery. *Nat. Nanotechnol.* **15**, 656–660 (2020).

71. Fornieri, A., Timossi, G., Virtanen, P., Solinas, P. & Giazotto, F. 0–π phase-controllable thermal Josephson junction. *Nature Nanotech* **12**, 425–429 (2017).

72. Glick, J. A. *et al.* Phase control in a spin-triplet SQUID. *Science Advances* **4**, eaat9457 (2018).

73. Hovhannisyan, R. A., Golod, T. & Krasnov, V. M. Controllable Manipulation of Semifluxon States in Phase-Shifted Josephson Junctions. *Phys. Rev. Lett.* **132**, 227001 (2024).

74. Tewari, S., Das Sarma, S., Nayak, C., Zhang, C. & Zoller, P. Quantum Computation using Vortices and Majorana Zero Modes of a $p_x+ip_y$ Superfluid of Fermionic Cold Atoms. *Phys. Rev. Lett.* **98**, 010506 (2007).

75. Fu, L. & Kane, C. L. Superconducting Proximity Effect and Majorana Fermions at the Surface of a Topological Insulator. *Physical Review Letters* **100**, 096407 (2008).

76. Elliott, S. R. & Franz, M. *Colloquium* : Majorana fermions in nuclear, particle, and solid-state physics. *Reviews of Modern Physics* **87**, 137–163 (2015).

77. Van Der Wal, C. H. *et al.* Quantum Superposition of Macroscopic Persistent-Current States. *Science* **290**, 773–777 (2000).

78. Clarke, J. & Wilhelm, F. K. Superconducting quantum bits. *Nature* **453**, 1031–1042 (2008).

## Acknowledgments

B.-C.L. acknowledges Tao Wu, Hongming Weng for valuable discussions. This work was supported by the National Key Research and Development Program of China (Grants No. 2022YFA1403700, No. 2020YFA0309300), the National Natural Science Foundation of China (Grant No. 12074162), Guangdong Basic and Applied Basic Research Foundation (Grant No. 2022B1515130005), Quantum Science and Technology-National Science and Technology Major Project (Grants No. 2021ZD0303000 and 2021ZD0303001). I.M. acknowledges financial support by the Swiss National Science Foundation (SNSF) via the SNSF postdoctoral Grant No. TMPFP2_217204. A. Sz. was supported by the Simons Foundation (Grant number SFI-MPS-NFS-00006741-11).

## Author contributions

B.-C.L. supervised the project, conceived the study, and designed the experiments. S.W. developed the etching technique. B.-C.L. and S.W. did the transport experiments. S. W. fabricated all devices in the main text. The supporting statistics experiments were done by S.W., Z.-N.W. and B.-C.L. J.-P.P., J.L. and Xuewen F. did the characterization experiments. Xilin. F., K.T.L., N.K., X.-S.W., D.Y., Y.X.Z., A.Sz., A.S. contributed to the analysis and interpretation of the experimental data. I.M., M.H.F. and M.S. developed the theoretical framework. B.-C.L., S.W., I. M., M.H.F. and M.S. wrote the manuscript with the necessary input from all authors.

## Competing interests

The authors declare that they have no competing interests.

## Figures legends

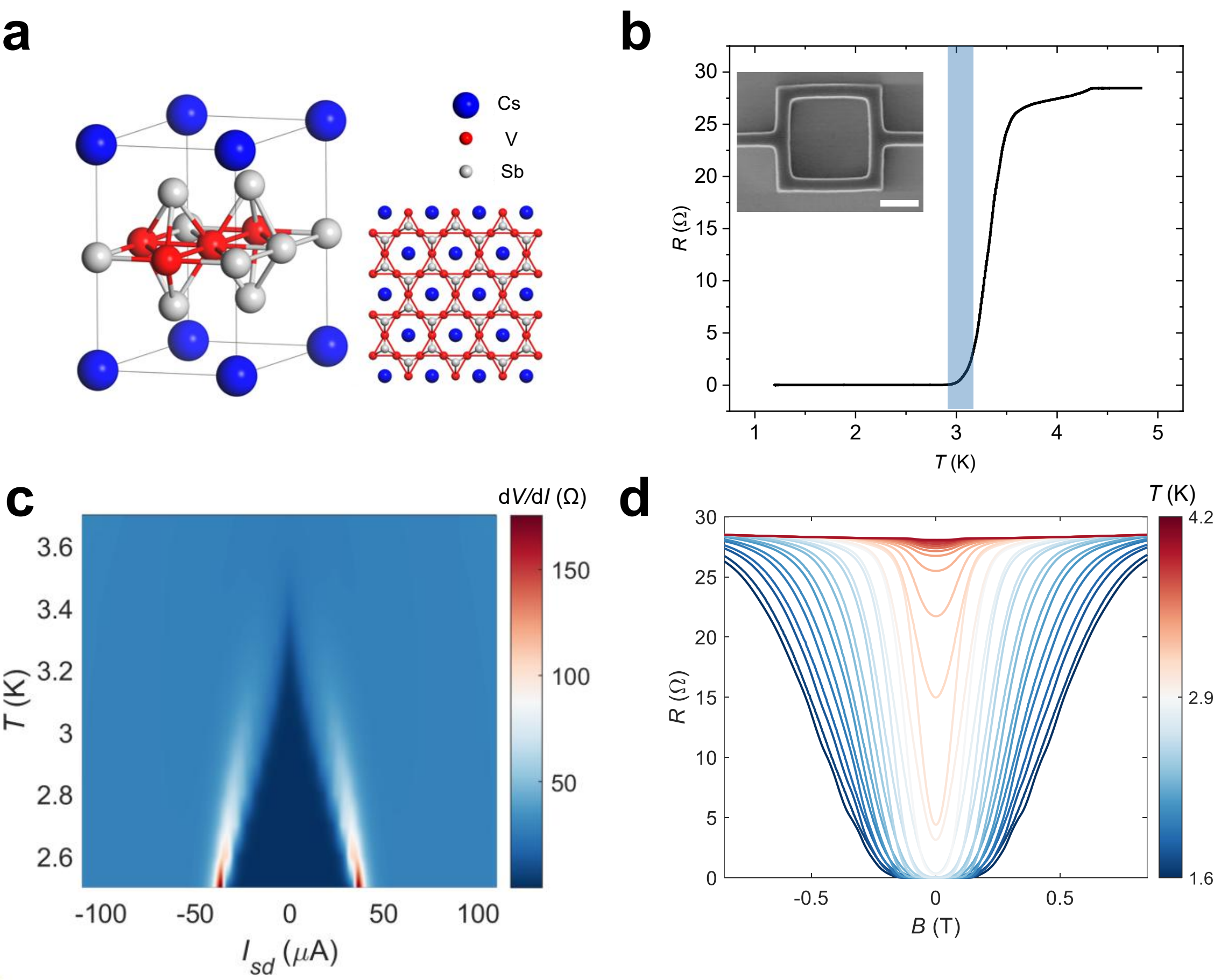

**Fig. 1 The superconducting properties of the mesoscopic ring of $CsV_3Sb_5$. a** The crystal structure of the $CsV_3Sb_5$ lattice. **b** The resistance-temperature curve of the $CsV_3Sb_5$ rings. The onset critical temperature of the superconductivity is around 2.9 K. The shallow blue region denotes where the π-phase Little-Parks oscillations can be observed, as will be shown below. Inset: The image of a typical device covered by photoresist, with a scale bar of 500 nm. **c** The colormap of differential resistance as a function of the bias current and temperature. The critical current decreases with the temperature increasing, showing a typical superconducting behavior. **d** The out-of-plane magneto-resistance as a function of temperature.

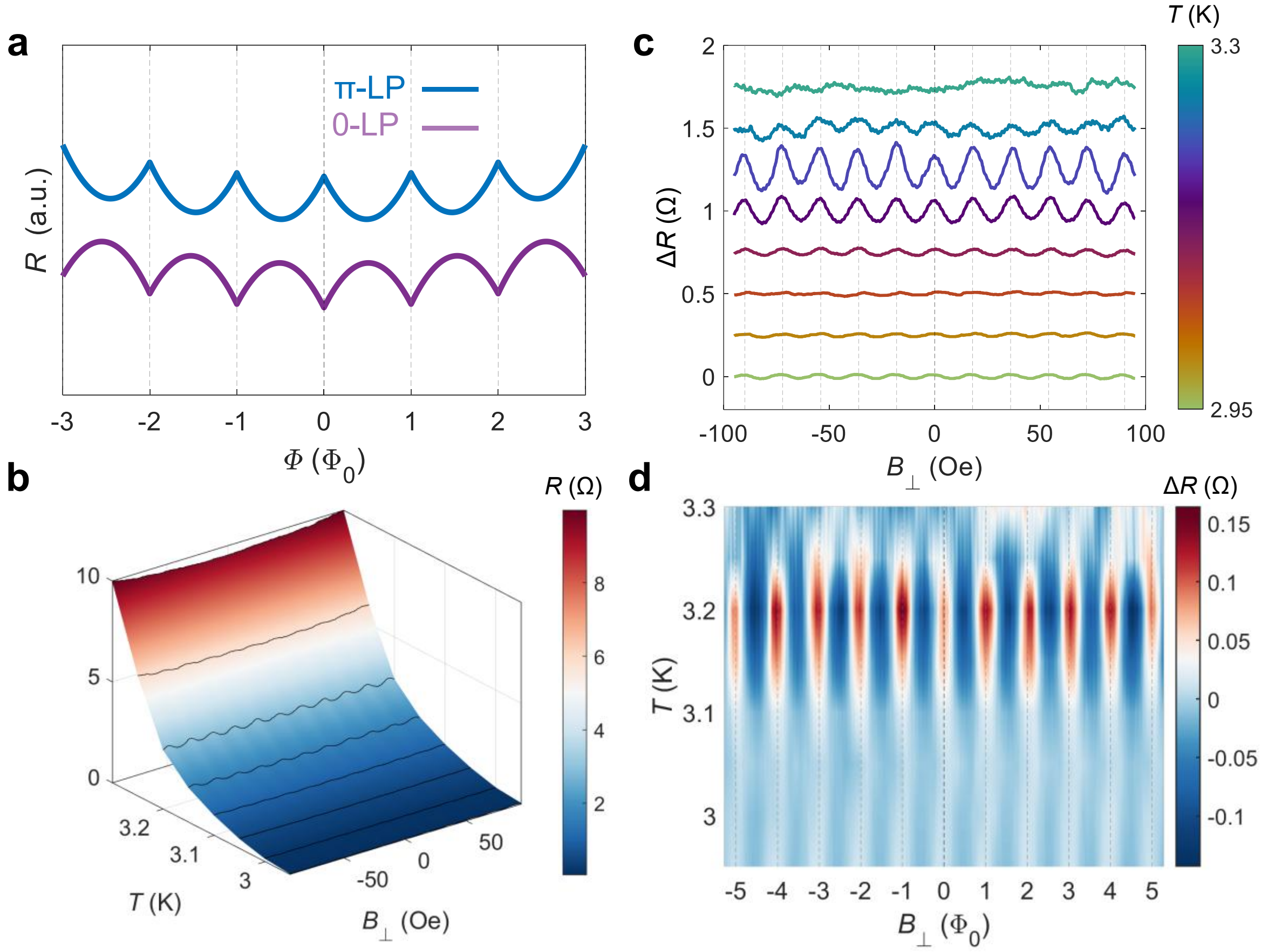

**Fig. 2 The phase transition of the π-phase Little-Parks effect driven by bias currents. a** The schematics of 0-phase and π-phase Little-Parks Oscillations. In conventional superconductors, the resistance minimum at $\Phi = n\Phi_0$ is observed near the critical temperature, while for unconventional superconductors with sign-changing superconducting order parameters, a π-phase shift can be introduced, resulting in quantization at $(n+1/2)\Phi_0$. **b** The original magnetoresistance oscillations of the mesoscopic rings as a function of temperature. Correspondingly, the extracted oscillations ΔR after subtracting the background resistance is displayed as a waterfall plot in **c** and a colormap in **d**. Here, the unconventional π-phase Little-Parks effect with resistance maximum at $\Phi = n\Phi_0$, in contrast to the conventional Little-Parks effect with resistance maximum at $\Phi = (n+1/2)\Phi_0$ is observed. In this sample, the period $\Phi_0$ is around 18 Oe.

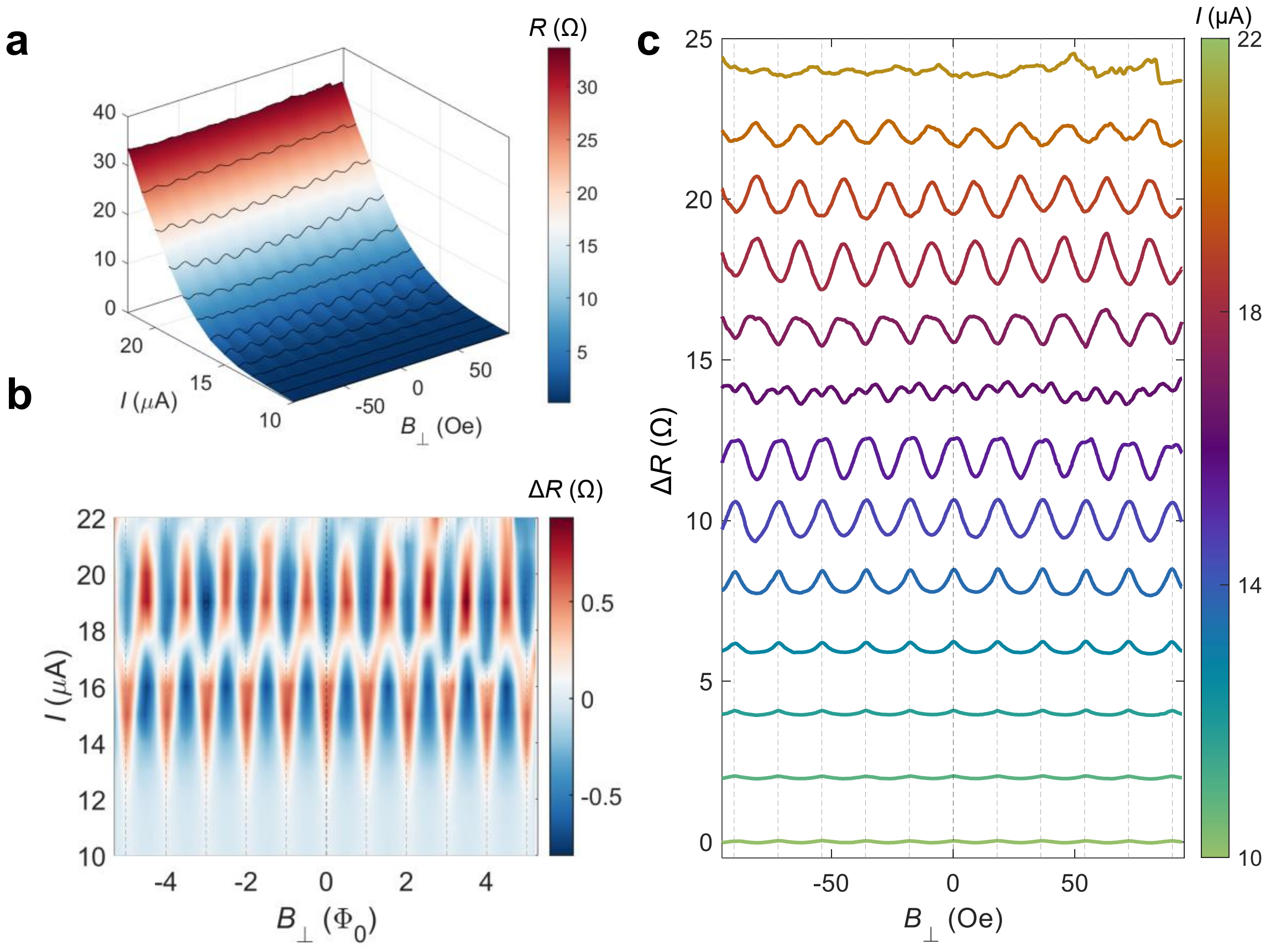

**Fig. 3 The phase transition of the π-phase Little-Parks to conventional Little-Parks effect driven by bias currents at T = 2.8 K. a** The colormap of the original magneto-resistance as a function of bias currents. At around 17 μA, the resistance peaks gradually transit to dips, indicating a phase transition. **b** The extracted oscillation resistance ΔR as a function of bias currents shows the phase transition more clearly. In other words, the unconventional π-phase Little-Parks effect with resistance minimum at $\Phi = (n + \frac{1}{2})\Phi_0$ transits to the conventional 0-phase Little-Parks effect with resistance minimum at $\Phi = n\Phi_0$. Here the x axis is in units of $\Phi_0$ = 18 Oe. **c** The waterfall plot of the extracted oscillation resistance ΔR as a function of bias currents. Here the axis is in units of Oe.

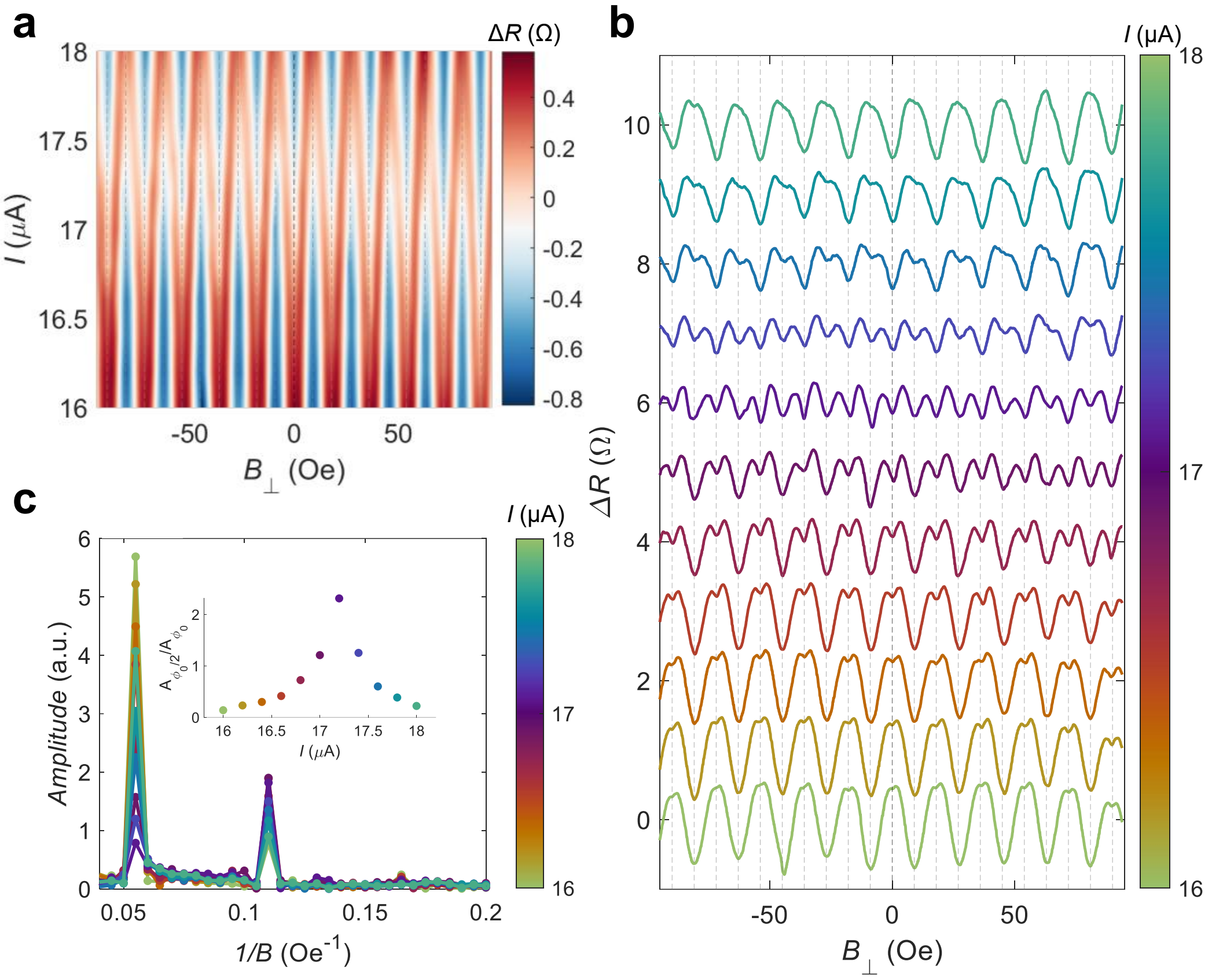

**Fig. 4 The evolution of the phase change and h/4e-period oscillations. a** The colormap of the extracted oscillations as a function of bias currents from 16 μA to 18 μA, which is the region where the phase transition happens. **b** The waterfall plot of the extracted oscillation resistance. During the phase transition, the half quantum flux period Little-Parks effect is observed, which is more pronounced at around 17.2 μA. **c** The fast-Fourier-transformation (FFT) analysis of the oscillating pattern during the phase transition. The FFT peaks at around quantum flux period and half quantum flux period. Inset: The ratio between the amplitude of half quantum flux period and quantum flux period $A_{\varphi 0/2}/A_{\varphi 0}$. Such ratio displays the dominance of the half quantum flux oscillating period, which peaks at 17.2 μA.

# Supplementary Information for

## Switchable half-quantum flux states in a ring of the kagome superconductor $CsV_3Sb_5$

*Corresponding Email:
mark.fischer@uzh.ch
sigrist@itp.phys.ethz.ch
linbenchuan@iqasz.cn

**The PDF file includes:**

## Section 2 Ginzburg-Landau theory of π-phase to 0-phase transition

# Section 1 The experiments

## Supplementary Note 1 How to find the π-phase Little-Parks effect in kagome superconductors?

1. High crystal quality of pristine $CsV_3Sb_5$ thin-flake microdevices.

Prior to etching process, thin flakes were mechanically exfoliated from bulk $CsV_3Sb_5$ crystals and carefully characterized. High-resolution transmission electron microscopy (HRTEM) image and selected diffraction (SAED) pattern image confirm the excellent crystallinity, as shown in Fig. S1a&b. The transport characterization further corroborates the high sample quality. As shown in Fig. S1c, the resistance vanishes below a superconducting transition temperature of $T_c$ = 3.1 K, slightly enhanced compared with bulk crystals, consistent with earlier reports on reduced dimensionality[1,2]. The in-plane upper critical field was determined from the midpoint criterion of the resistive transition (50% of the normal resistance), yielding $B_c \sim 8.8$ T (Fig. S1d) at the lowest temperature 0.1 K. This value substantially exceeds the weak-coupling Pauli paramagnetic limit $B_p = 1.86T_c = 5.766$ T, giving $B_c/B_p = 1.53$. Such a clear violation of the Pauli limit highlights possible unconventional nature of superconductivity in $CsV_3Sb_5$ and at the same time, attests to the pristine quality of our thin flakes, since reliable measurement of Pauli-limit violation requires exceptionally good sample quality.

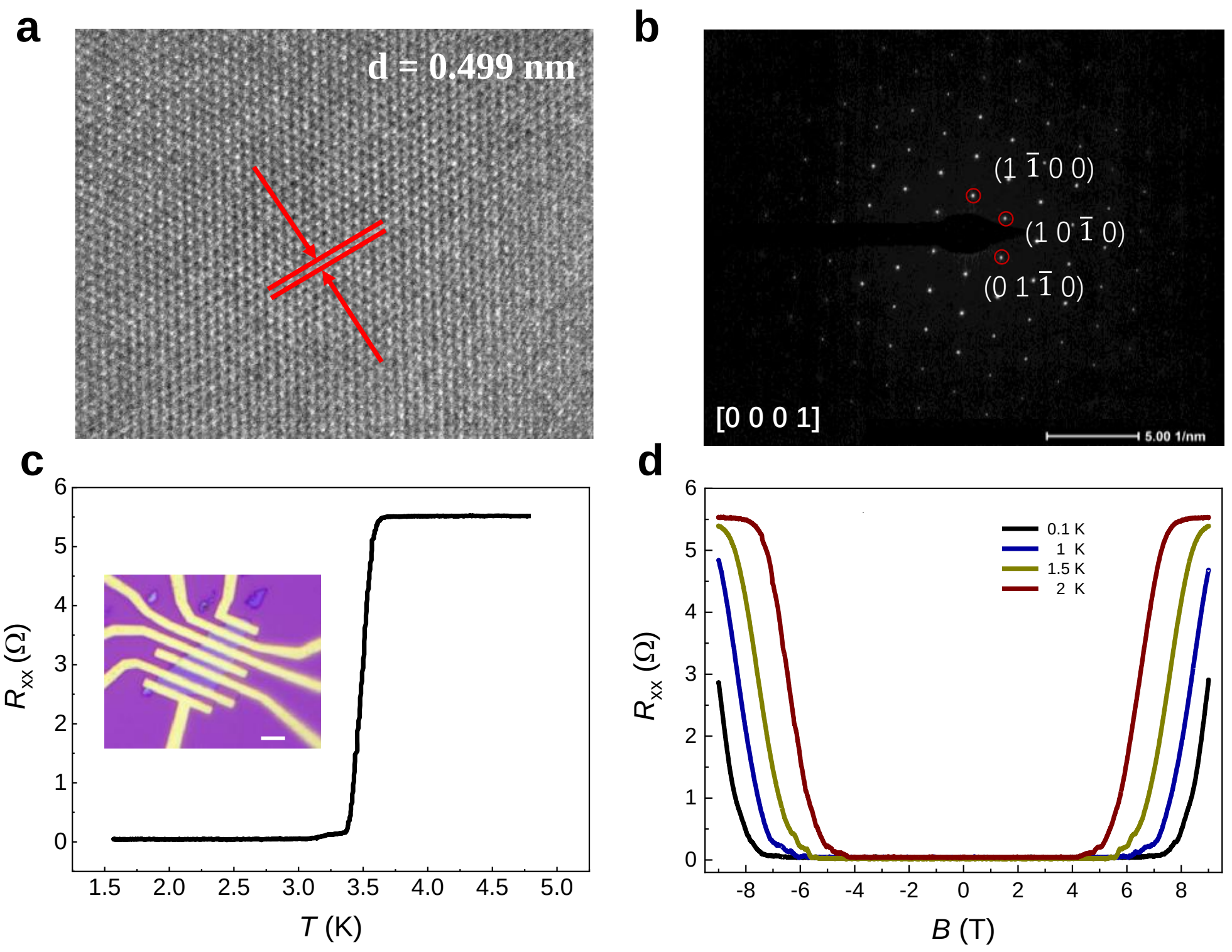

**Fig. S1 The characterization of pristine $CsV_3Sb_5$ thin flakes. a** The HRTEM image of the cleaved sample. **b** The SAED image of the sample. **c** The transfer curve of the mechanically exfoliated thin flakes and its optical image. The scale bar is 3 μm. Generally, the device is encapsulated by hBN to protect the sample from oxidation. **d** The magnetoresistance curve under in-plane magnetic fields at different temperatures.

2. Fabrication technique.

Superconducting ring structures provide a powerful platform to probe the phase information of the condensate wavefunction, as demonstrated in systems such as $Sr_2RuO_4$[3], $Bi_2Pd$[4–6], and $TaS_2$[7,8]. Various microfabrication approaches have been employed to define similar mesoscopic geometries. Among them, focused ion beam (FIB) miling[9,10] and reactive ion etching (RIE)[11] are the two most common techniques, with inductively coupled plasma (ICP) etching regarded as an advanced variant of RIE. Because ICP and RIE share essentially the same working principle[11], we restrict the present discussion to RIE and FIB.

The two methods differ fundamentally. FIB employs a highly focused beam of energetic ions to physically sputter atoms from the sample surface, enabling direct patterning but often inducing significant subsurface disorder[9,10]. In contrast, RIE is a plasma-based process in which chemically reactive radicals (F, Cl, O, etc.) etch the sample surface through ion-assisted reactions localized near the sample sheath. As a result, RIE generally produces less structural damage in crystalline superconductors[11], since ion penetration is shallow and largely confined to the surface. Nevertheless, the actual extent of damage not only depends on the etching methods (RIE or FIB), but also depends sensitively on the etching conditions, recipe optimization[9–11], and the tolerance of the specific material system. Therefore, both RIE and FIB etching have therefore been widely and successfully adopted in the fabrication of superconducting microdevices.

From the device perspective, the extent of etching-induced damage can be evaluated by monitoring changes in the transport characteristics. In particular, abrupt anomalies in the resistance-temperature (R-T) curve or an overly broad R-T transition would indicate substantial disorder. As shown in Fig. 1a, the superconducting transition temperature defined by the deviation from the zero-resistance state is $T_c = 2.9$ K, only slightly reduced compared with the pristine sample ($T_c = 3.1$ K, Fig. S1c). This minimal suppression demonstrates that our loop device retains nearly the same superconducting quality as the original crystal. *Such parameter comparison provides a direct benchmark for assessing whether the chosen etching technique is optimal for given material.* In principle, similar comparisons can be performed across different fabrication methods on the same material system. Maintaining such pristine characteristics is particularly critical in kagome superconductors, where fragile band topology and correlation effects can be obscured by disorder or extrinsic perturbations[12]. Consistently, Fig. S2 shows that the etched loop device (sample #1) sustains a large in-plane critical field ~ 6.2 T, still comparable to the Pauli-limit violation in pristine samples, further confirming the integrity of our microfabrication process.

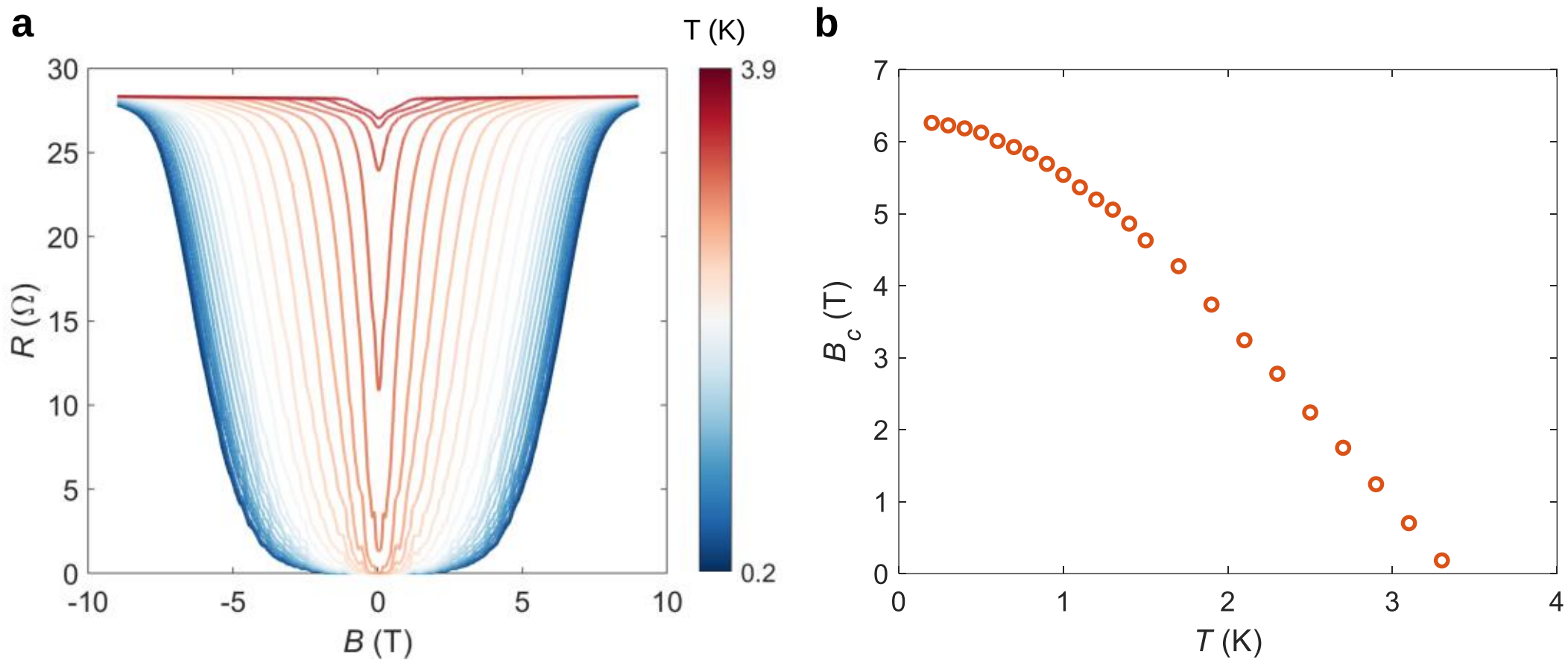

**Fig. S2** The magnetoresistance curves of ring-structure $CsV_3Sb_5$ thin flakes under in-plane magnetic fields at different temperatures, as shown in **a**. The extraction of the critical field is shown in **b**. The critical field is still comparable to the field of Pauli-limit violation.

3. Little-Parks oscillations.

In their seminal 1962 work[13], Little and Parks demonstrated charge-2e oscillations in superconducting cylinders, providing direct evidence of magnetic flux quantization. Several experimental conditions are important when reproducing such measurements. First, the oscillations can be detected either by monitoring the resistance or the transition temperature; resistance measurements are generally more straightforward and widely used. Second, reliable observations require working in the low-field regime, since high fields not only introduce additional flux-related or band-structure effects but also suffer from technical complications such as flux trapping and an ill-defined zero-field point. Below we will discuss many technique issues involved.

a) Low-field regime: In our study, all measurements were carefully performed within $\pm 100$Oe, ensuring that the remnant field of the superconducting magnet remained negligible. The magnitude of this remnant field typically scales with the maximum field previously applied; for example, after sweeping to 1 T, a residual field of ~20 Oe or more can persist. By contrast, within the $\pm 100$ Oe range, the residual field is reduced below 1 Oe and can be suppressed to $\sim 0.1$ Oe with sufficiently slow field sweeps.

b) Sweep rate: The sweep rate itself is another critical factor. If the magnetic field is ramped too quickly, the forward and backward sweeps exhibit a finite offset. Slower sweeps, in contrast, yield nearly overlapping traces, effectively canceling such hysteresis and allowing the zero-field point to be determined with high accuracy. For instance, as shown in Fig. S3, at a sweep rate of 6 Oe/min, the near-zero oscillation peak occurs at +1.77 Oe in the forward sweep and at -1.69 Oe in the backward sweep, giving an average zero point of 0.04 Oe. At a slower rate of 3 Oe/min, the offsets reduce further, the near-zero oscillation peak occurs at +0.25 Oe in the forward sweep gives 0.25 Oe and at -0.2 Oe in the backward sweep, yielding an average zero point of 0.025Oe. This procedure not only helps to choose the appropriate sweep rate but also enables us to calibrate the absolute zero field with an accuracy better than 0.04 Oe, thereby ensuring that the Little-Parks oscillations are measured around a well-defined zero-field reference.

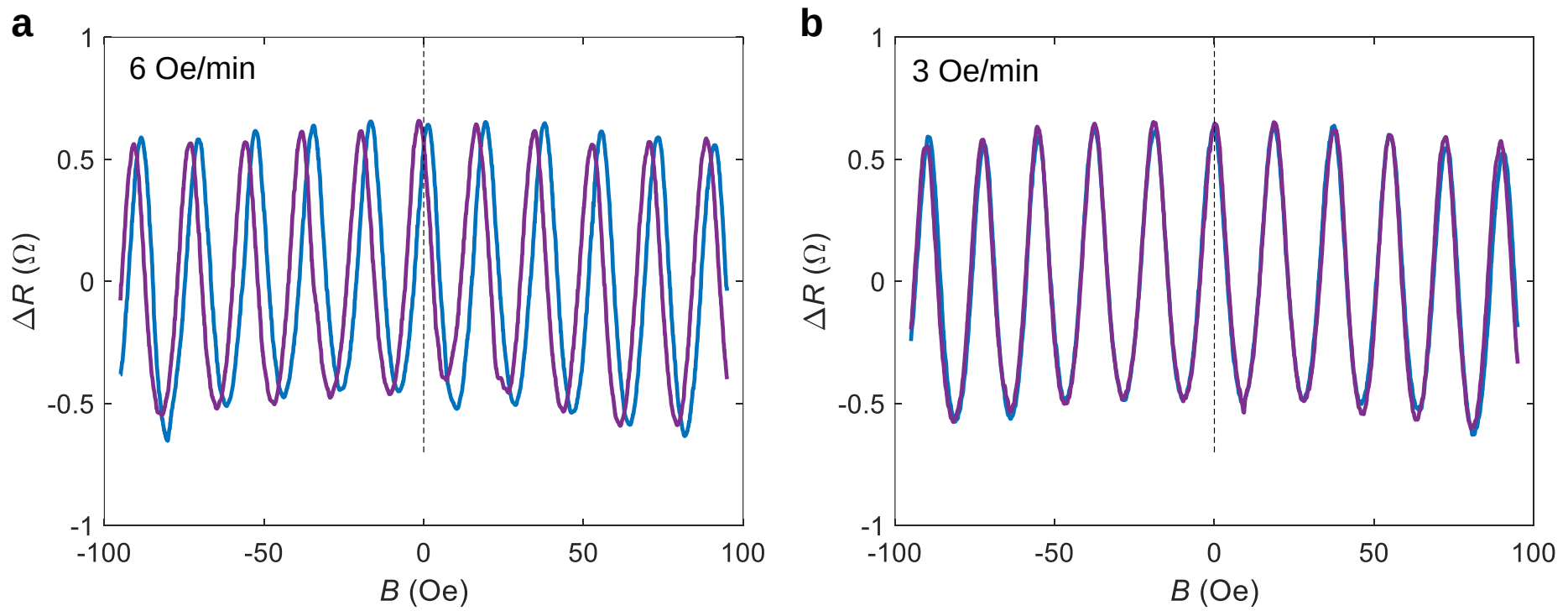

**Fig. S3** The π-phase Little-Parks oscillations under different sweeping rates of magnetic fields.

c) High-temperature resistance curve: An additional method to determine the zero-field point is to use the high-temperature magnetoresistance curve, where oscillations are absent. In this regime, the magnetoresistance is expected to be symmetric with respect to magnetic field, and the position of the resistance minimum therefore corresponds to zero field (Fig. S4). It should be noted, however, this approach is limited in accuracy, with an uncertainty on the order of 1 Oe in our case.

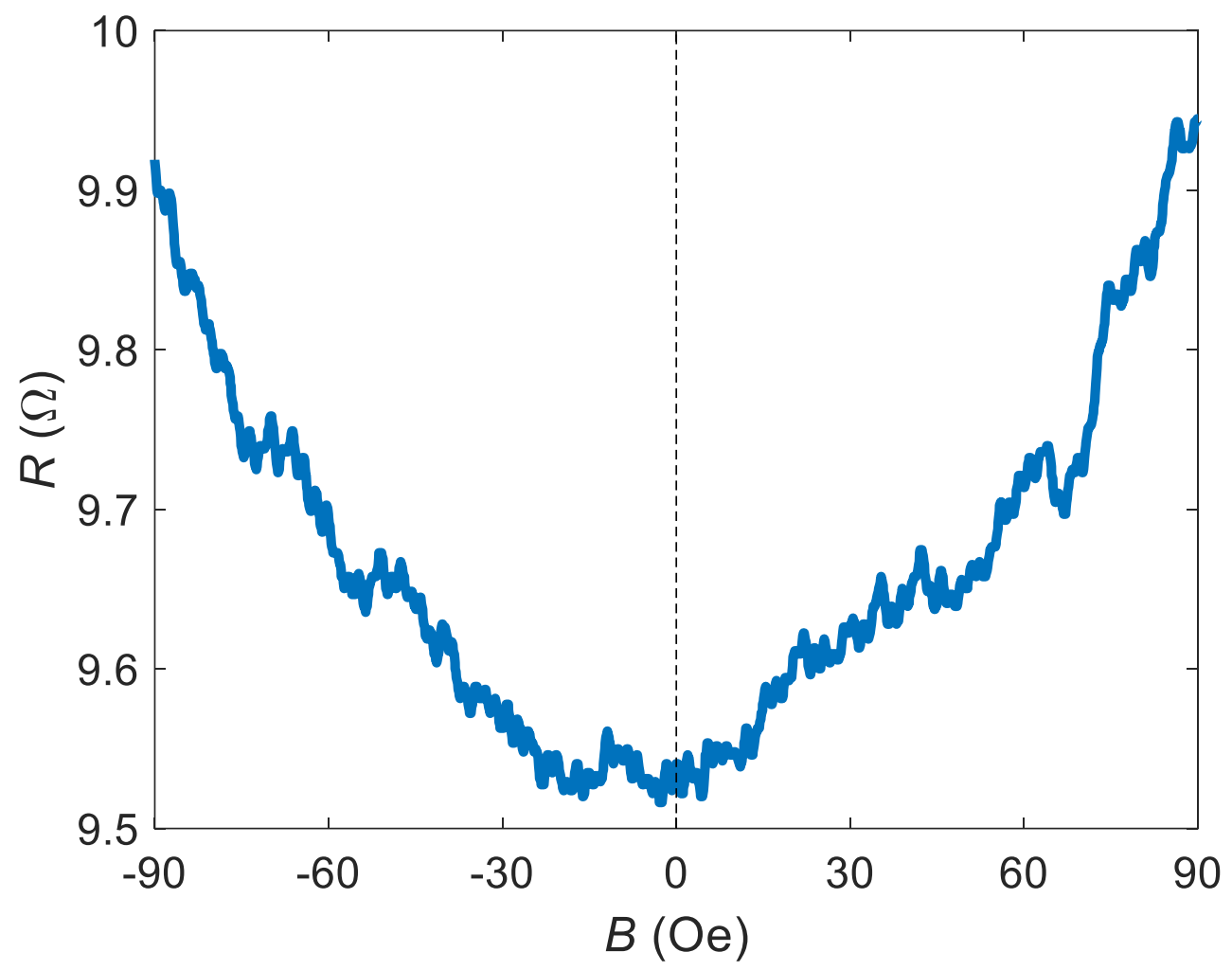

**Fig. S4** The magnetoresistance curve at 3.3 K of sample #1.

d) Loop size and oscillation period: Despite the careful procedures outlined above, a residual remnant magnetic field of order 0.04 Oe may still persist due to experimental limitations. To minimize its influence, the loop dimensions were chosen such that the oscillation period is much larger than this residual field. In our devices, the loop has an average area of ~ 1 μm$^2$, corresponding to an oscillation period at the order of ~ 20 Oe—nearly three orders of magnitude greater than the estimated remnant field. A larger period also mitigates the effect of self-induced fields from the bias current: for instance, a current of 1 μA generates an effective field ~ 0.01 Oe, far smaller than the oscillation period observed here.

e) Thermal-cycle test

Thermal cycling provides an additional verification of the stability of the oscillations, and a means to detect possible flux trapping in the magnet or the sample. In this procedure, the device was first warmed above $T_c$ to the normal state, during which the superconducting magnet was quenched to release any trapped flux. The sample was then re-cooled to 2.8 K. As shown in Fig. S5, the π-phase Little-Parks oscillations, the transition between π and 0 phases, and the intermediate h/4e oscillations all reappear without detectable phase shifts. These results confirm that the observed oscillations are intrinsic and remain stable against thermal cycling.

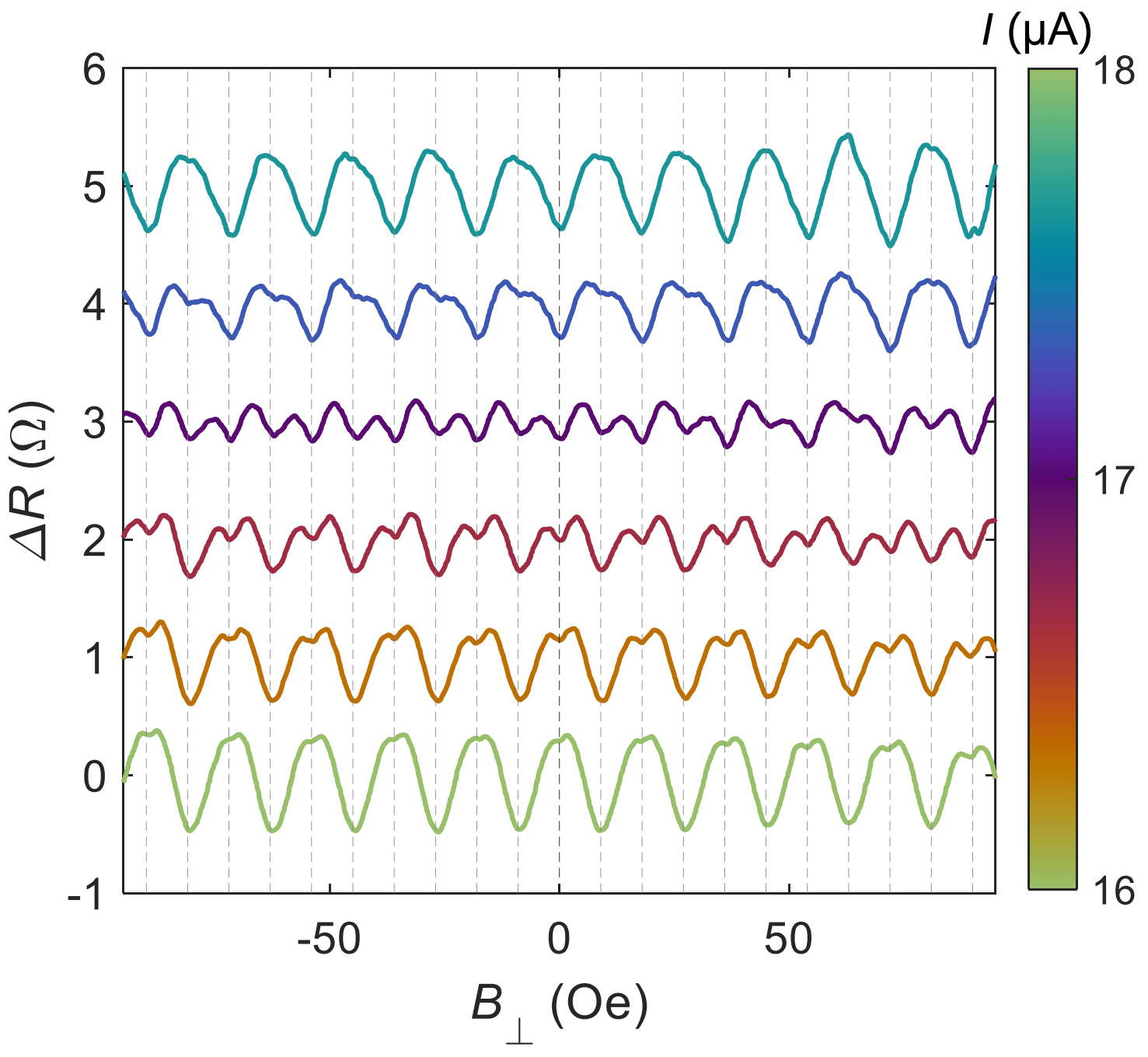

**Fig. S5 The oscillations remain independent of the thermal cycle**. During the thermal cycle, both the sample and superconducting magnet are heated over 10 K to enter into the normal states and then cool the sample again to the superconducting states. The oscillations themselves and the evolution driven by bias currents remain the same.

f) Physical mechanisms: From the perspective of physical mechanisms, trivial vortex-related effects can be ruled out as the origin of the observed π-phase Little-Parks oscillations. If the π phase were caused by flux pinning or similar vortex physics, it would be difficult to account for the evolution from the π-phase to the 0-phase oscillations that we observe. Likewise, scenarios based on unbalanced current distribution in the loop can be excluded. In that case, it would generate a continuous phase shift that tracks the bias-current amplitude, resulting in a gradual π-to-0 crossover rather than the discrete phase switching (π, 0, or h/4e-periodicity) revealed in our measurements.

## Supplementary Note 2 Reproducibility and statistics

Demonstrating the π-phase Little-Parks effect is experimentally challenging, and reliable conclusions cannot rest on a single device, particularly for loop structures fabricated without external junctions. For polycrystalline or strongly inhomogeneous samples, prior work has invoked a Josephson weak-link scenario, in which grain boundaries act as Josephson junctions[4]. In such cases, the loop may acquire an odd or even number of sign changes depending on the random distribution of grains, leading to a statistical 50:50 distribution of π- and 0-phase oscillations. Beyond this extrinsic explanation, other mechanisms have been proposed for single-crystalline systems without grains or junctions, including nonuniform spin-triplet d-vector textures[14,15], strain-field effects[8] and negative Josephson couplings between neighboring layers[16].

In our study, we systematically investigated more than 10 devices. Strikingly, almost all of them except one (sample #13) showed the π-phase effect, as shown in Fig. S6~S8. The origin of this single 0-phase device remains unclear. It was observed only during the early stage of this project and subsequent fabrication efforts have not succeeded in reproducing the original 0-phase behavior in later devices. Taken together, our statistics strongly indicate that the π-phase Little-Parks oscillation is the dominant and reproducible outcome in high-quality kagome superconducting loops.

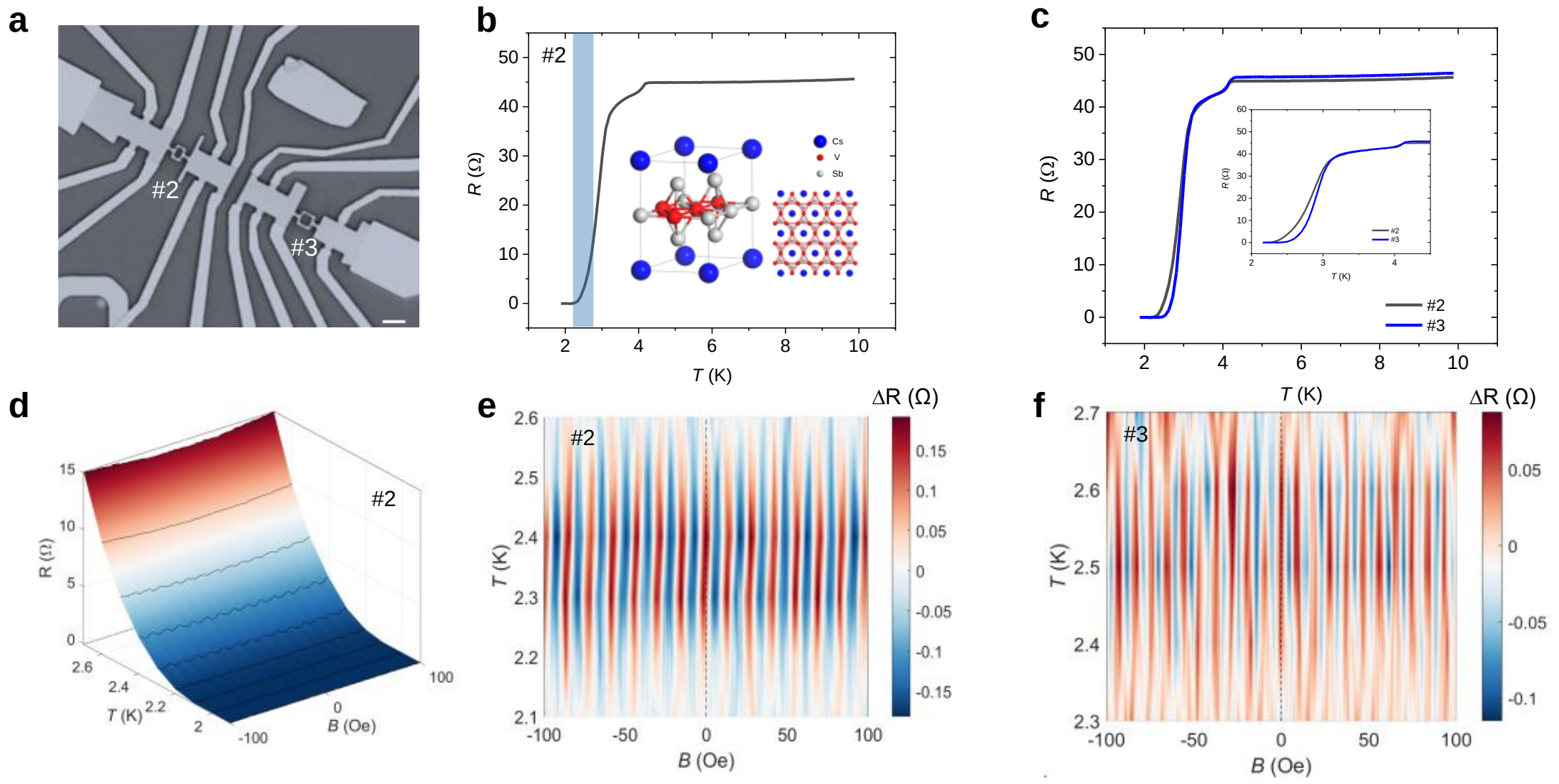

**Fig. S6 The π-phase Little-Parks oscillations in samples #2 and #3. a** The image of device #2 and #3, both etched from the same pristine flake. The scale bar is 2 μm. The loop areas are approximately 1.5 μm$^2$ and 2.25 μm$^2$ for samples #2 and #3. **b** Resistance-temperature curve of sample #2. The blue-shaded region indicates the temperature range where π-phase Little-Parks oscillations are observed (see **d**, **e**). **c** Resistance-temperature curves of sample #2 and #3. **d** Raw magnetoresistance curves of sample #2 at different temperatures. **e** Extracted oscillations of sample #2 at different temperatures of sample #2, revealing clear π-phase Little-Parks periodicity. **f** Extracted oscillations of sample #3, showing more periods of π-phase Little-Parks effect between ±100 Oe, consistent with the relatively larger loop area.

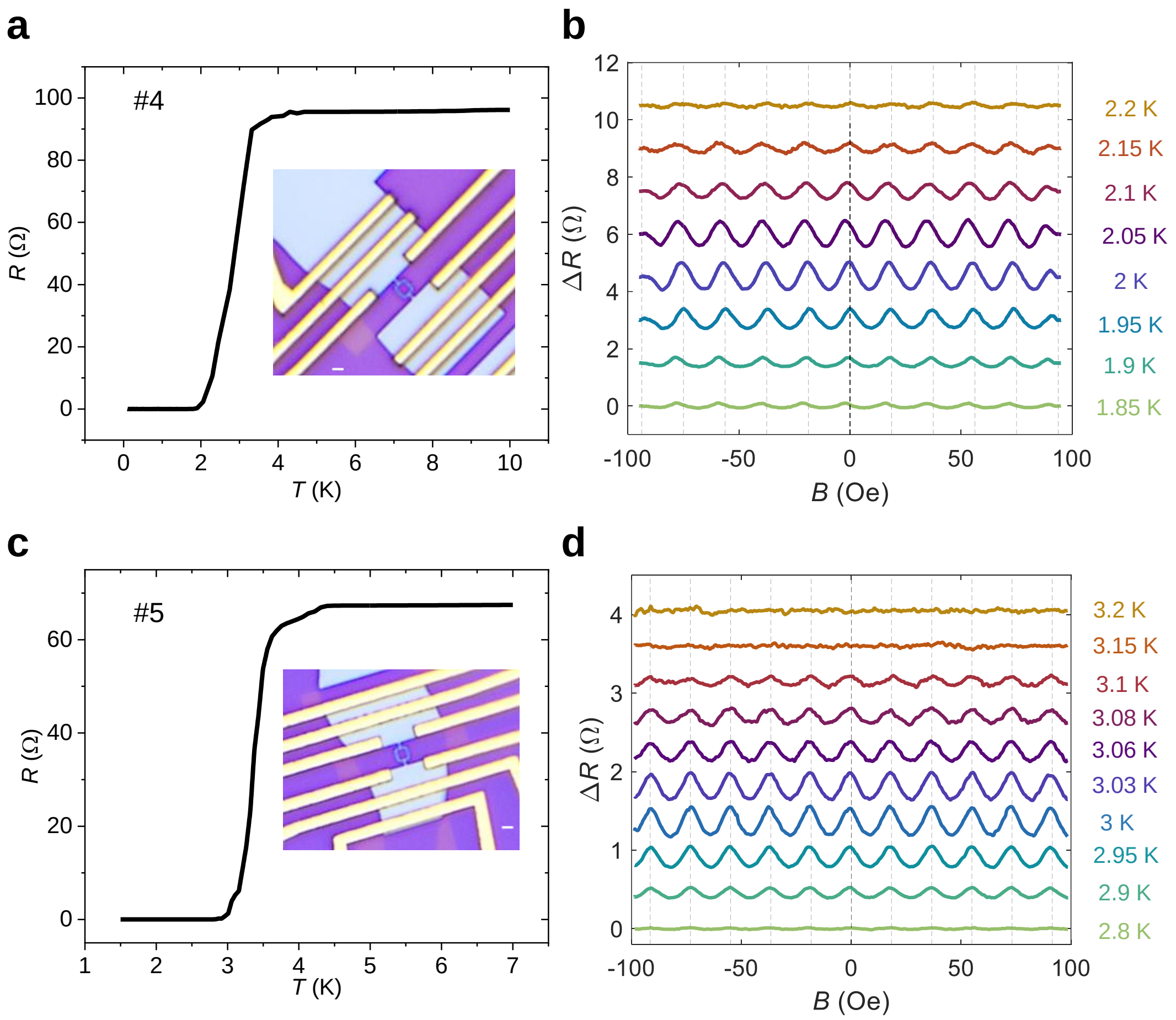

**Fig. S7 The π-phase Little-Parks oscillations in samples #4 and #5. a,c** The resistance-temperature curves of device #4 and #5. The scale bar is 1 μm. **b,d** The extracted oscillations of sample #4 and #5 at different temperatures, showing π-phase Little-Parks effect.

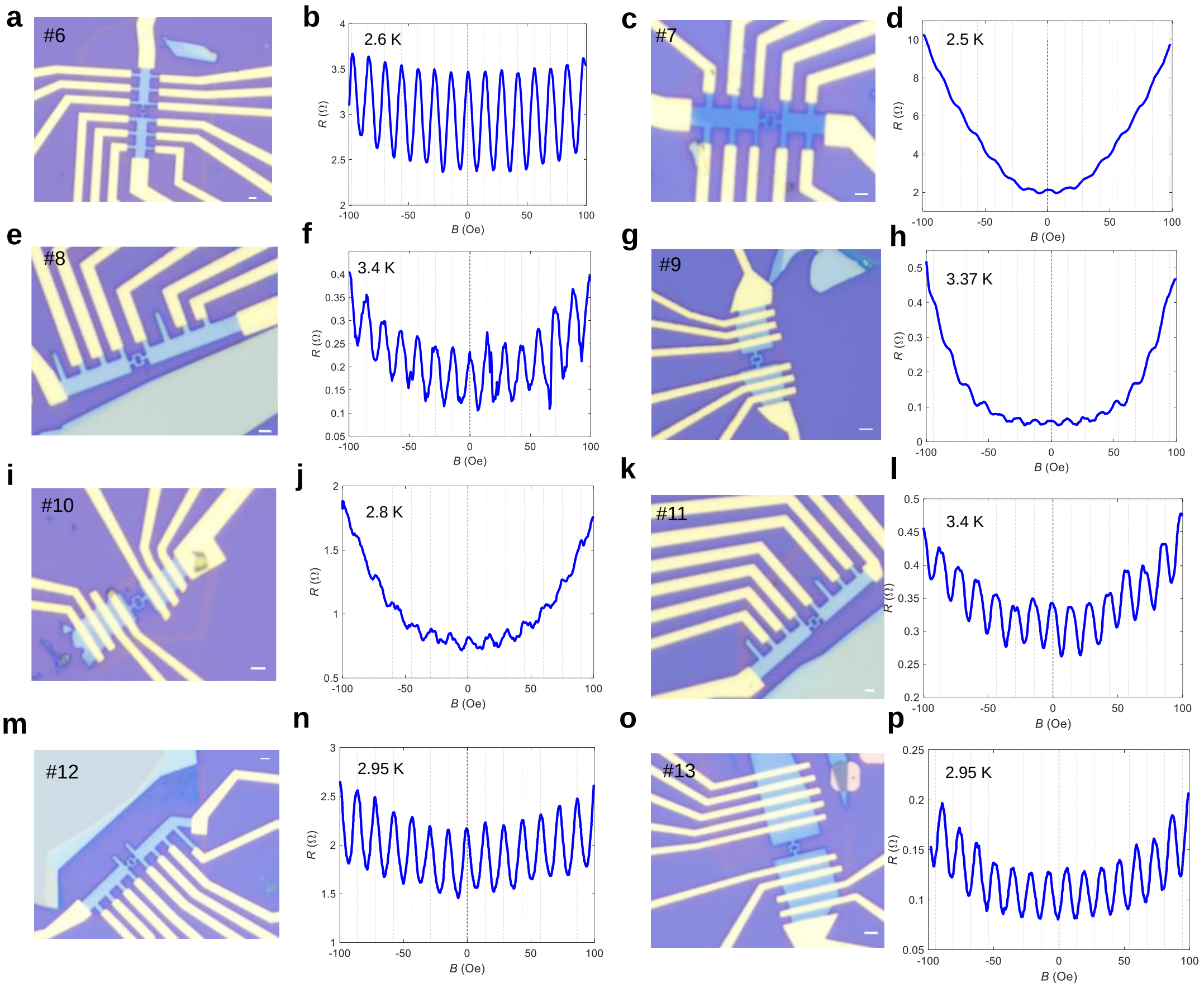

**Fig. S8 Statistics of Little-Parks oscillations in samples #6 ~ #13.** The optical images and corresponding magnetoresistance oscillations are shown. The scale bar is 2 μm. Across different devices, the overall quality and critical temperature are different, leading to differences in the optimal temperature range for observing the Little-Parks oscillations. Despite these variations, clear Little-Parks oscillations were consistently observed. Except for sample #13, all other devices exhibit π-phase Little-Parks oscillations. The 0-phase LP oscillations observed in sample #13 has not been reproducibly confirmed. In other words, nearly all measured devices intrinsically display the π-phase Little-Parks effect.

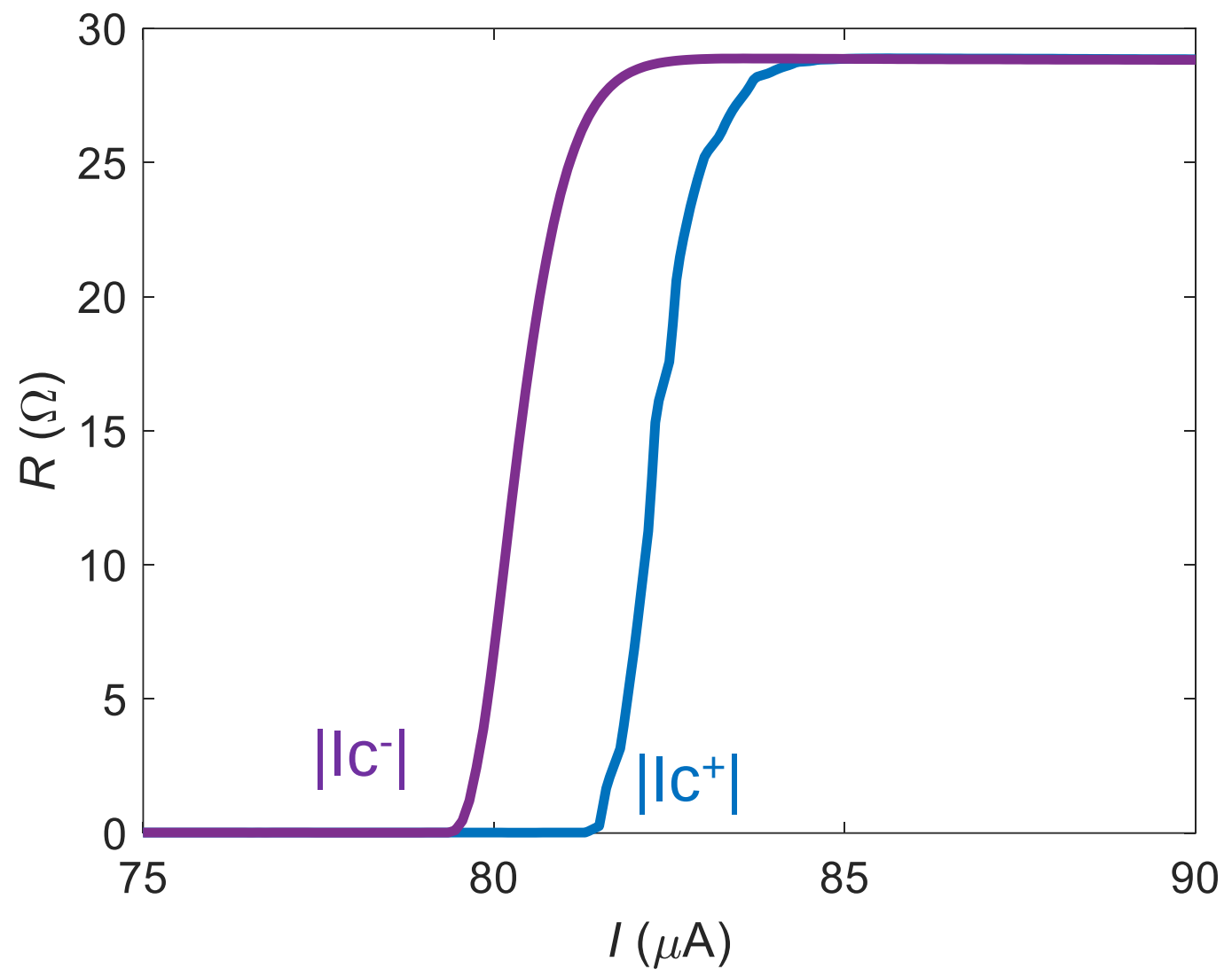

**Fig. S9 Zero-field superconducting diode effect observed in the $CsV_3Sb_5$ ring**. The temperature is 20 mK.

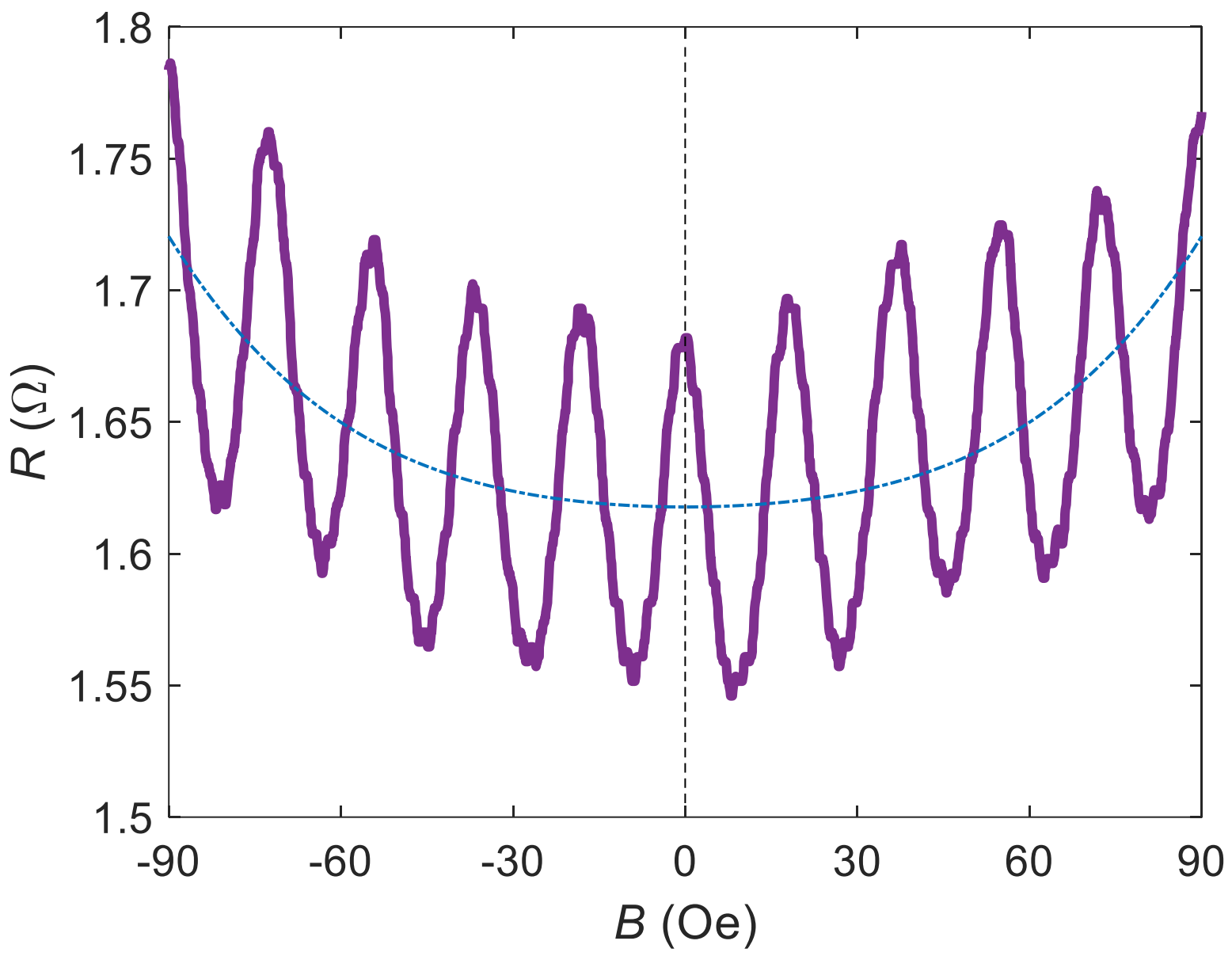

**Fig. S10 The background extraction of the original magneto-resistance curve.** Here the background signal is fitted via the polynomial even function.

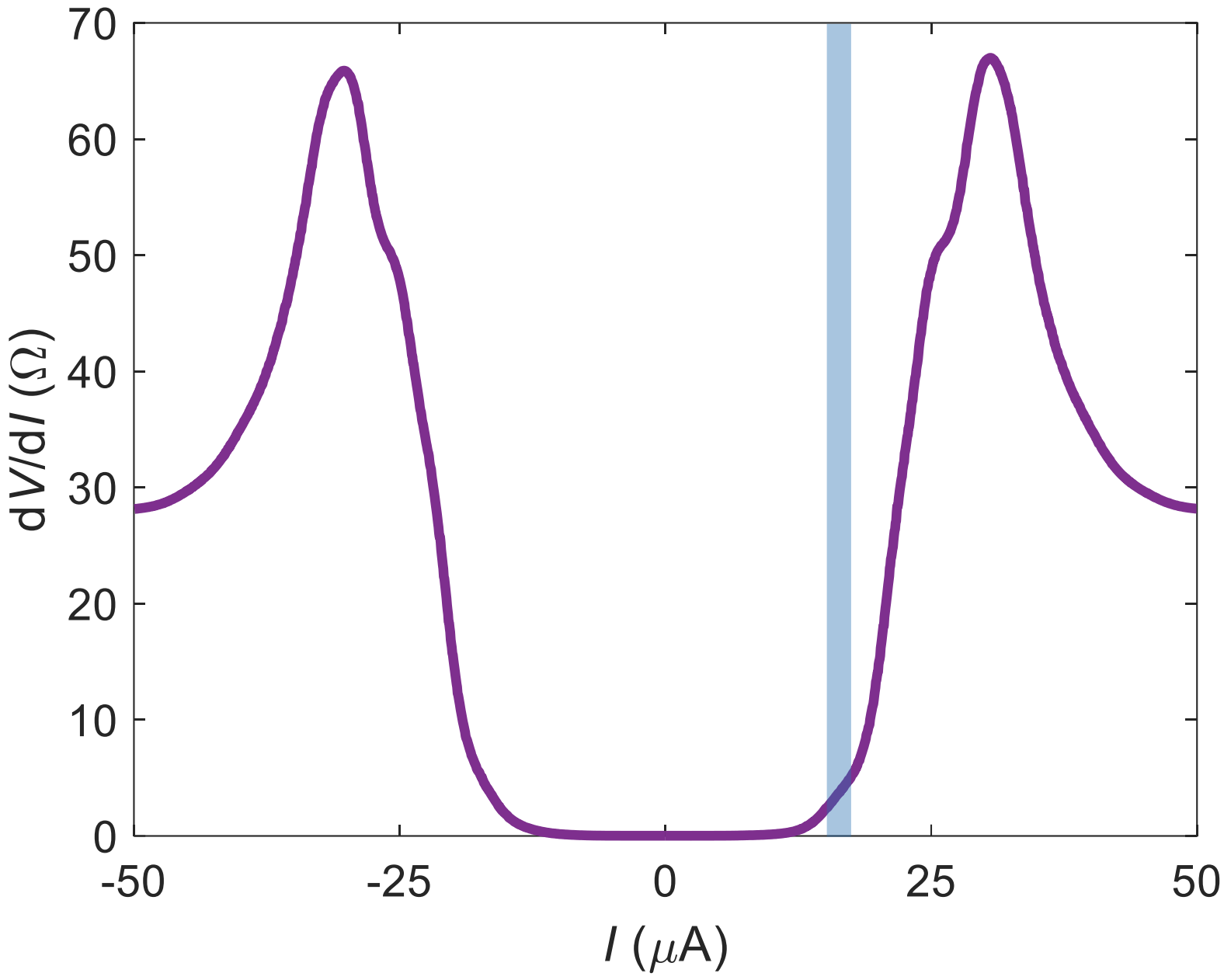

**Fig. S11 The region where the current-driven phase evolution happens in sample #1.**

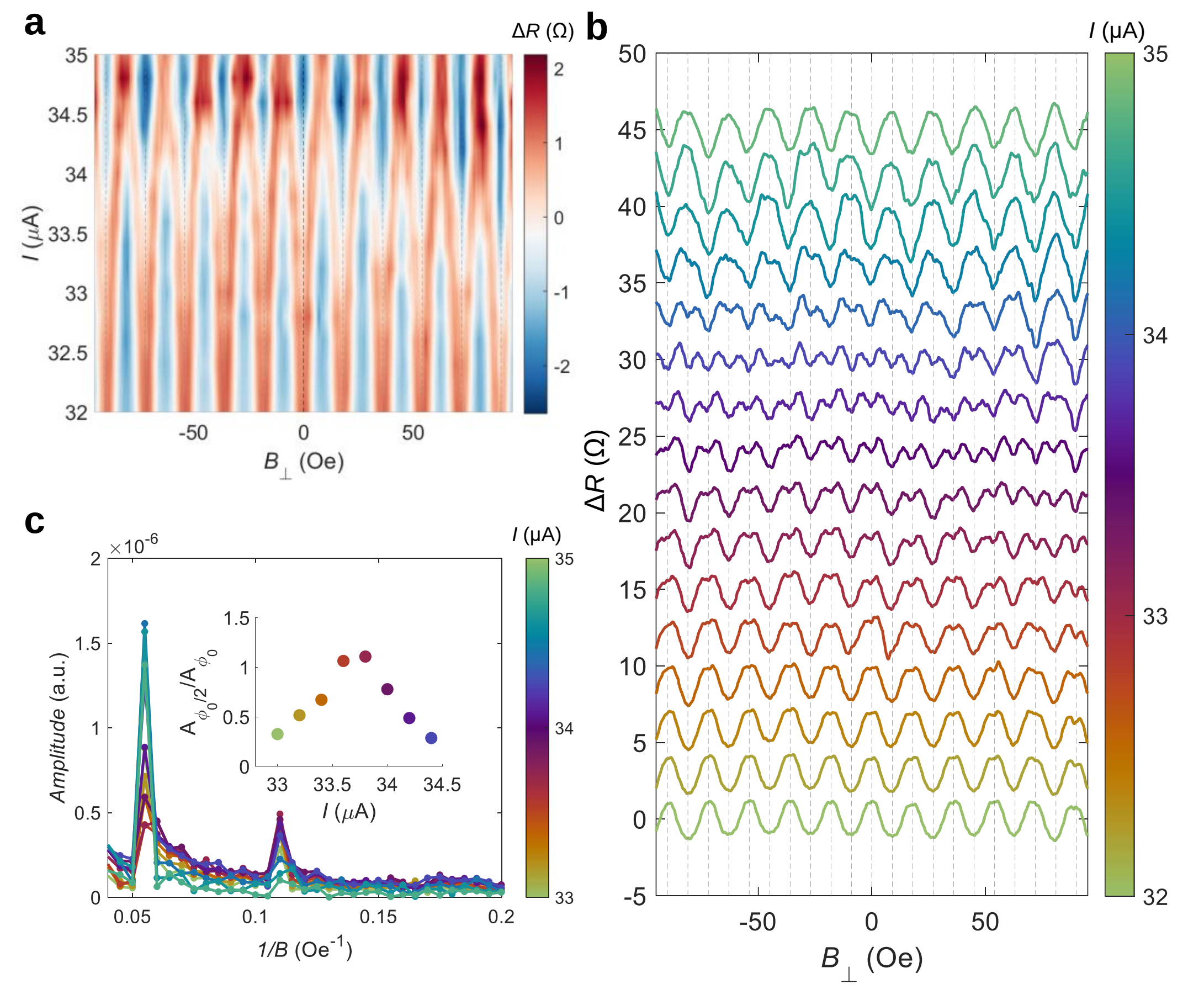

**Fig. S12** The phase evolution between π-phase and 0-phase LP effects and h/4e oscillations at another temperature T = 2.5 K of sample #1.

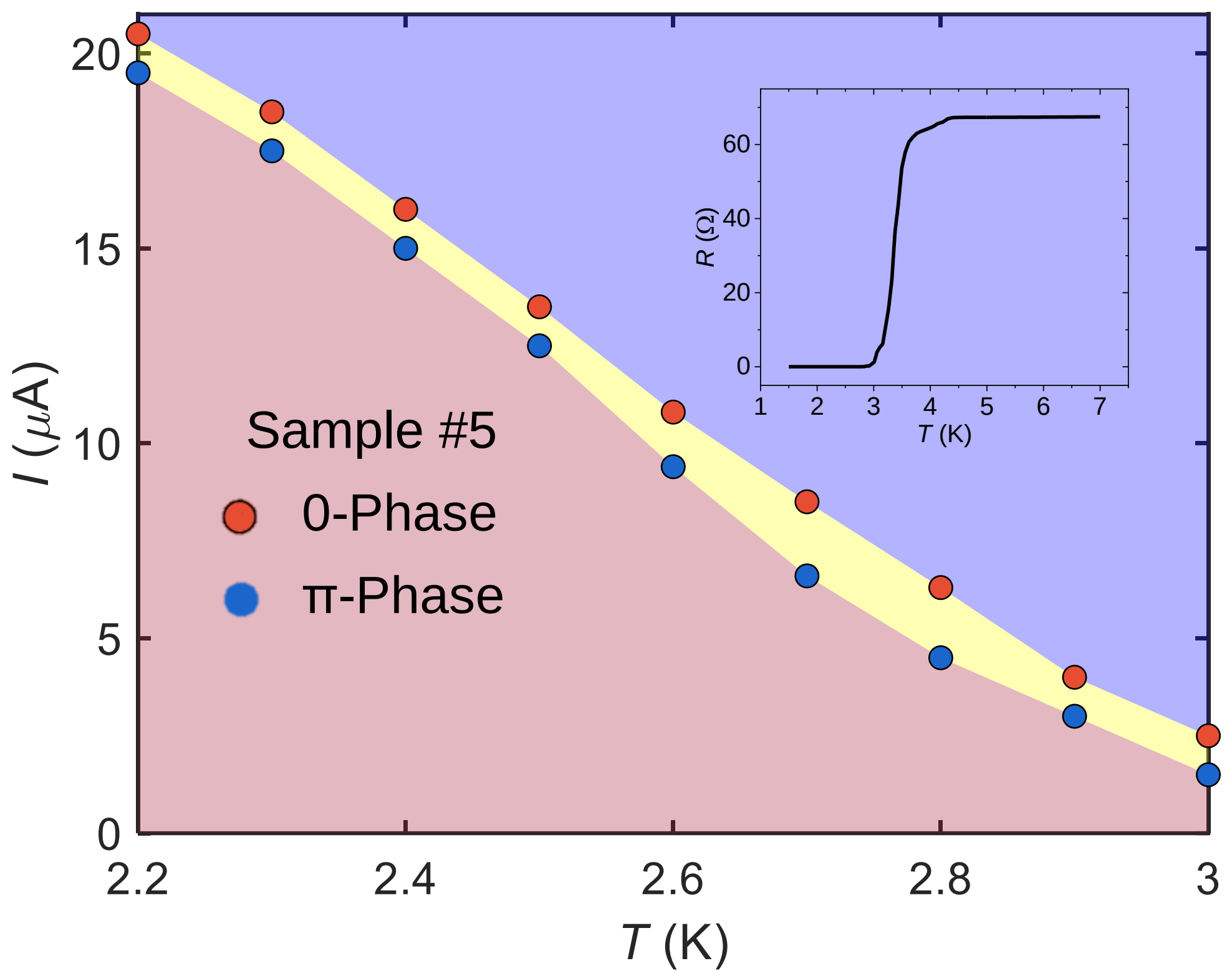

**Fig. S13** The phase diagram of another sample #5 showing the π-phase and 0-phase, between which the h/4e oscillations emerge.

**Supplementary Note 3 The background signals of current-driven LP oscillations**

For the temperature-driven Little-Parks Oscillations, the background was extracted using a polynomial even-function fit ($R= R_2*B^2+R_0$), which yields a background symmetric to the magnetic field, as shown in Fig. 2 and Fig. S8. In contrast, for the current-driven Little-Parks oscillations, the raw magnetoresistance data exhibit a pronounced asymmetry with the magnetic field. In Figs. 3,4 of the main manuscript, to better isolate the oscillatory components, the background was therefore removed using a general polynomial fit ($R = R_2*B^2 + R_1*B^1+ R_0$) to focus on the oscillations.

In this supplementary note, we will discuss the background signals. As shown in Fig. S14, if using the even-function polynomial fit ($R =R_2*B^2 + R_0$), besides the oscillations, there is a residual linear background, particularly at higher bias currents 16 μA to 20 μA.

When a negative current is applied, the background signal reverses its polarity, as shown in Fig. S15, indicating that the asymmetry depends jointly on both the magnetic field and current directions. Figs. S15a,b further reveals a subtle phase shift between positive and negative currents, possibly arising from the superconducting diode effect as shown in Fig. S9 and the associated time-reversal symmetry breaking. Moreover, as shown in Fig. S16, the background signals of intermediate h/4e period oscillations also exhibit a clear dependence on current direction.

In summary, the π-phase Little-Parks oscillations have an intrinsic symmetric background with the magnetic field. However, when a bias current is applied, the asymmetric background signal with the magnetic field is involved. Such asymmetry depends both on the magnetic field and bias current direction, potentially reflecting a structural or electronic imbalance between the two loop arms.

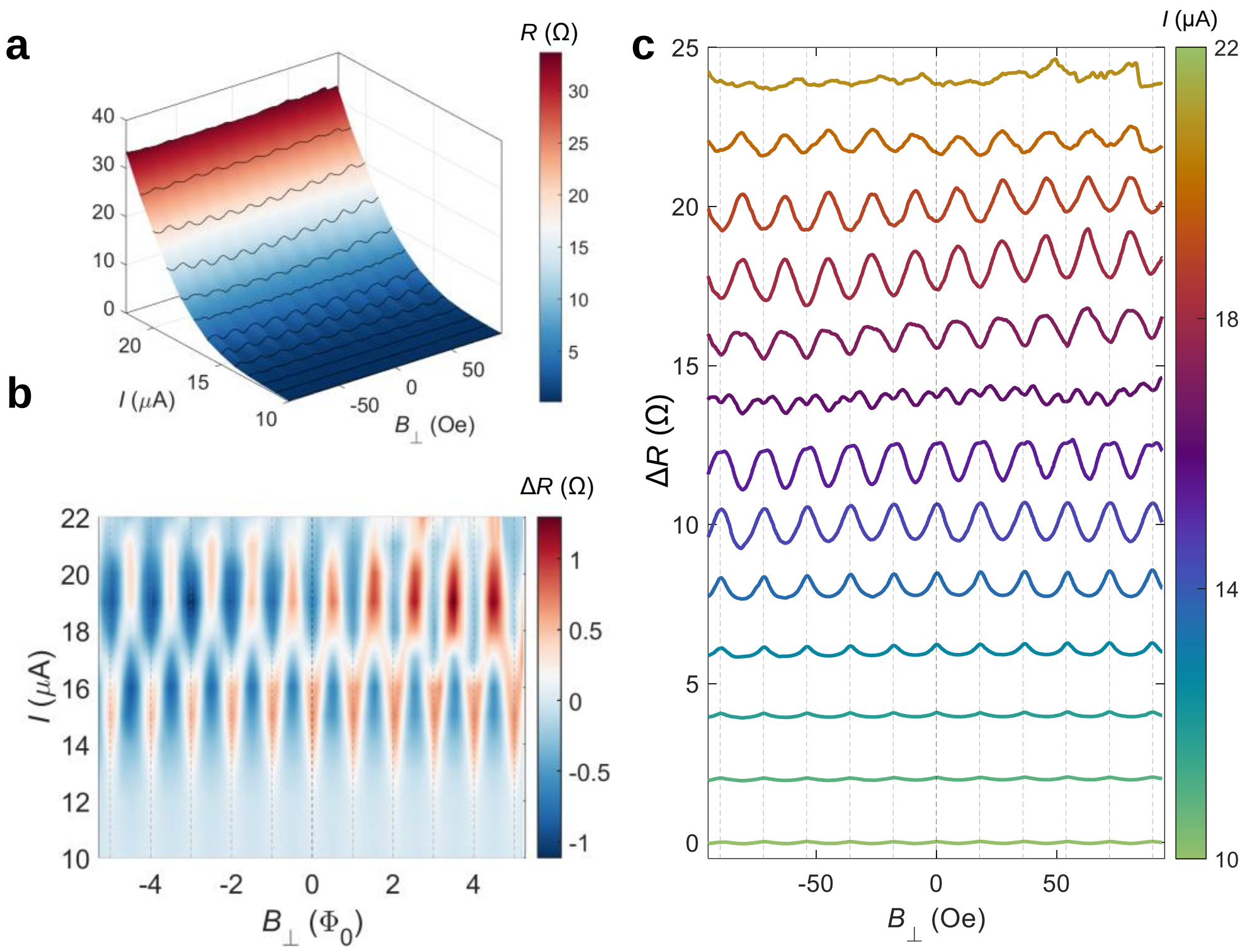

**Fig. S14 Comparison of background extraction methods for current-biased Little-Parks oscillations.** Same data as in Fig. 3 but analyzed using a different background-removal approach. In the main text, the background signal was extracted according to conventional polynomial fit ($R = R_2 * B^2 + R_1 * B^1 + R_0$) to emphasize the oscillatory component. Here the background signal was extracted according to an even-function polynomial fit ($R = R_2 * B^2 + R_0$). Thus, the asymmetric background components with respect to the magnetic field was demonstrated in **b** and **c**.

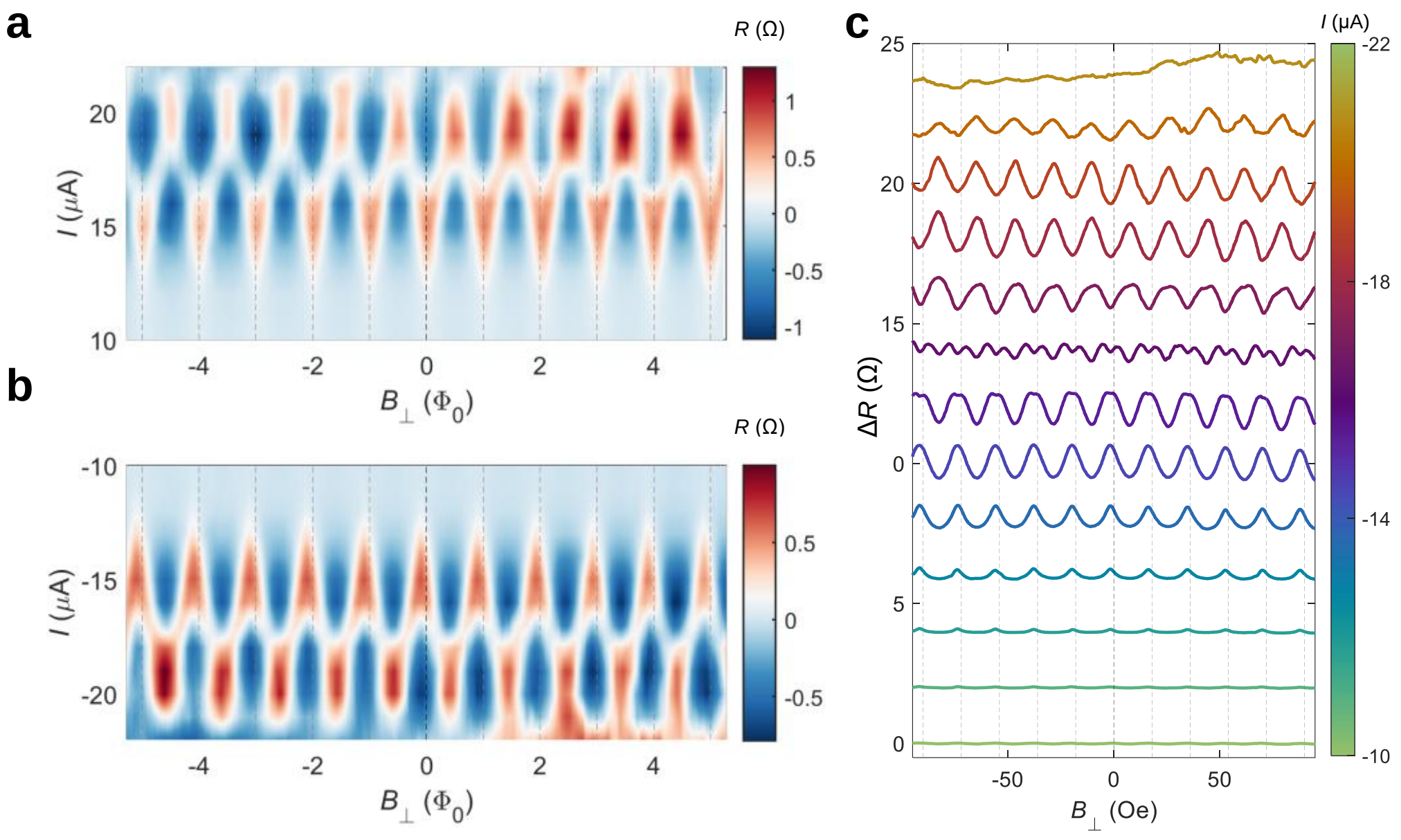

**Fig. S15 Current-driven Little-Parks oscillations under opposite current directions.** The background signal was extracted using an even-function polynomial fit. **a**,**b** There is a minor phase shift under negative current bias. The background signals exhibit opposite slopes for opposite current directions. **c** the waterfall plot of the Little-Parks oscillations under negative currents.

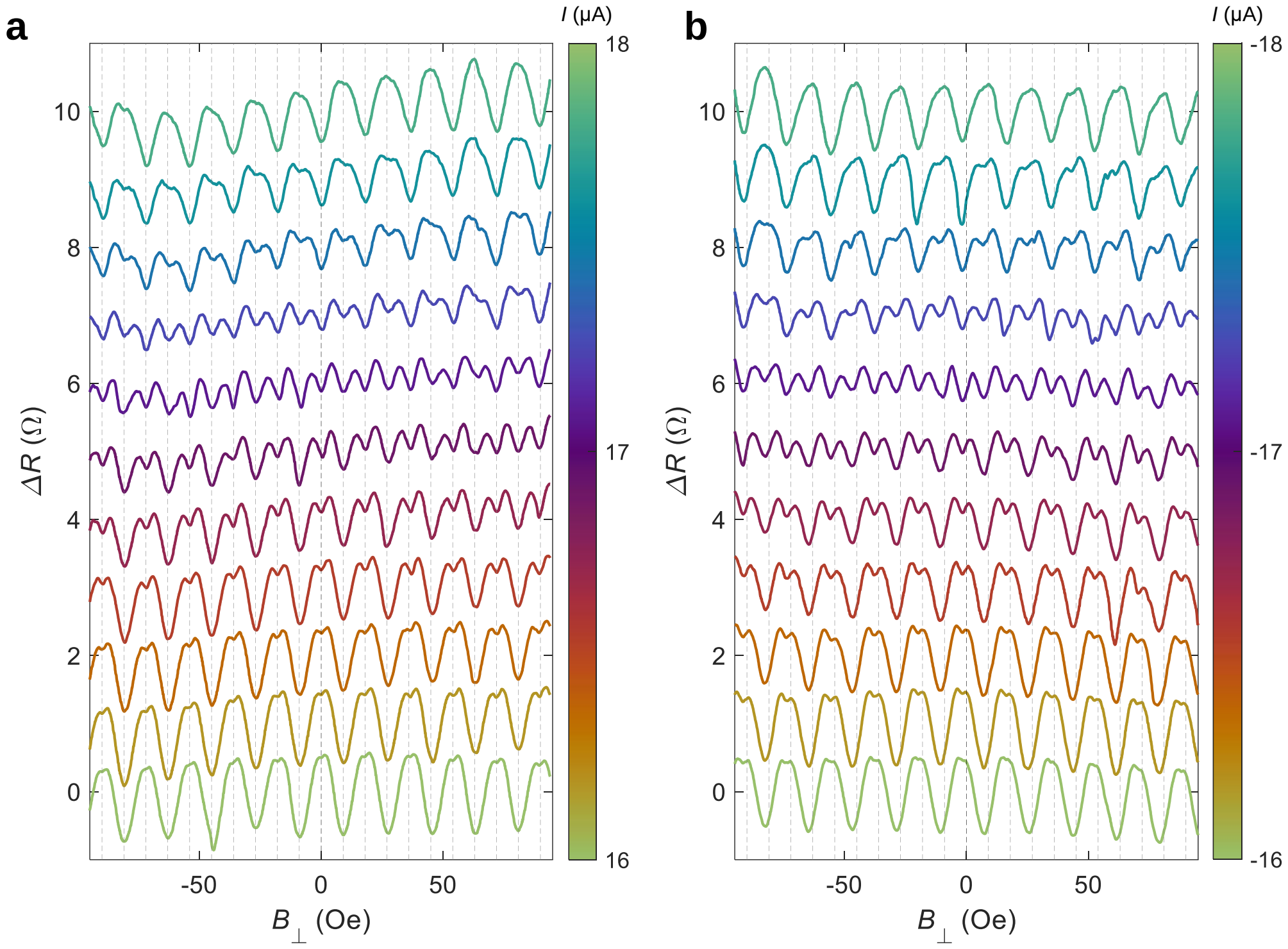

**Fig. S16 The h/4e-period oscillations with opposite current directions.** The background was extracted according to the even-function polynomial fit, same as Fig. S15. The bias currents are 16 μA to 18 μA in **a** and -16 μA to -18 μA in **b**.

## Section 2 Ginzburg-Landau theory of π-phase to 0-phase transition

Here, we provide an illustrative model to explain how a current can induce a switch from the π-phase to 0-phase Little-Parks effect. More detailed models require an understanding of the microscopic mechanism leading to a systematic π-phase ring at zero and low currents, aspects that remain to be explored.

The observed switch in current from a π-ring to a 0-ring is most straight-forwardly described using more than one superconducting component. To simplify the discussion, we assume one dominant and one subdominant superconducting component. These could be order parameters discussed along the lines of Refs.[8,16].

At low currents, almost all the measured samples display a π-phase shift of the Little-Parks effect suggesting the presence of some intrinsic mechanism favoring a superconducting component with a π-phase winding around the ring. Without making any assumption on the origin of the π-phase winding, we label $\psi_\pi$ the dominant superconducting component at zero and low bias current and $\psi_0$ the subdominant superconducting component stabilized at finite bias current. The Ginzburg-Landau free-energy density for the two components reads

$$f_{0,\pi} = \alpha_{0,\pi}|\psi_{0,\pi}|^2 + \frac{\beta_{0,\pi}}{2}|\psi_{0,\pi}|^4 + K_{0,\pi}|\vec{D}\psi_{0,\pi}|^2$$

with $\vec{D} = \vec{\nabla} - ie\vec{A}$ , $\alpha_{0,\pi} = a_{0,\pi}(T - {T_{c_0}}^{0,\pi})$ and ${T_{c_0}}^{0,\pi}$ is the bare mean-field critical temperature of the two components. For simplicity, we neglect any direct coupling between the two order parameters, which would be allowed in fourth order. To account for the ring geometry, we rewrite the free energy in cylindrical coordinates. To simplify the discussion, we consider only the azimuthal angle $\theta$, writing the covariant derivative as $D_\theta = \frac{1}{R}\partial_\theta - ieA_\theta$, while keeping fixed the radial coordinate $R$. Finally, by assuming that the superconducting components have uniform amplitude along the ring, we only consider the (integer) winding of their phase. With these assumptions, by integrating the free-energy density around the ring, we find

$$f_\pi = \alpha_\pi|\psi_\pi|^2 + \frac{\beta_\pi}{2}|\psi_\pi|^4 + \frac{K_\pi}{R^2}|\psi_\pi|^2\left(n + \frac{1}{2} - \frac{\Phi}{\Phi_0}\right)^2$$

for the component having a π-phase winding around the ring, and

$$f_0 = \alpha_0|\psi_0|^2 + \frac{\beta_0}{2}|\psi_0|^4 + \frac{K_0}{R^2}|\psi_0|^2\left(n - \frac{\Phi}{\Phi_0}\right)^2$$

for the subdominant component, with $n$ an integer. The two critical temperatures as a function of the applied magnetic field are the respective maximal critical temperatures and read

$$T_c{}^{\pi} = \max_n\left[T_{c_0}{}^{\pi} - \frac{K_\pi}{a_\pi R^2}\left(n + \frac{1}{2} - \frac{\Phi}{\Phi_0}\right)^2\right];$$

$$T_c{}^{0} = \max_n\left[T_{c_0}{}^{0} - \frac{K_0}{a_0 R^2}\left(n - \frac{\Phi}{\Phi_0}\right)^2\right].$$

The experimental evidences suggest that at low current $T_c{}^{\pi} > T_c{}^{0}$. To account for the effect of a finite current, we begin by discussing the standard Little-Parks effect from a different perspective, i.e. by computing the external flux dependence of the critical current[17]. The circulating current density along the ring reads

$$J_\theta = \frac{2eK}{R^2}|\psi|^2\left(n - \frac{\Phi}{\Phi_0}\right) = e|\psi|^2 v_\theta,$$

where for simplicity we omit the superconducting component indices and $v_{\theta\ 0,\pi} = \frac{2K}{R^2}\left(n_{0,\pi} - \frac{\Phi}{\Phi_0}\right)$, with $n_0 = n;\ n_\pi = n + \frac{1}{2}$, is the supercurrent velocity. As a function of the external magnetic flux $\Phi$, the superconducting system will minimize the value of the supercurrent velocity $v_\theta$, via the winding number of its phase, i.e.

$$v_\theta = \min_n\left[\frac{2K}{R^2}\left(n - \frac{\Phi}{\Phi_0}\right)\right].$$

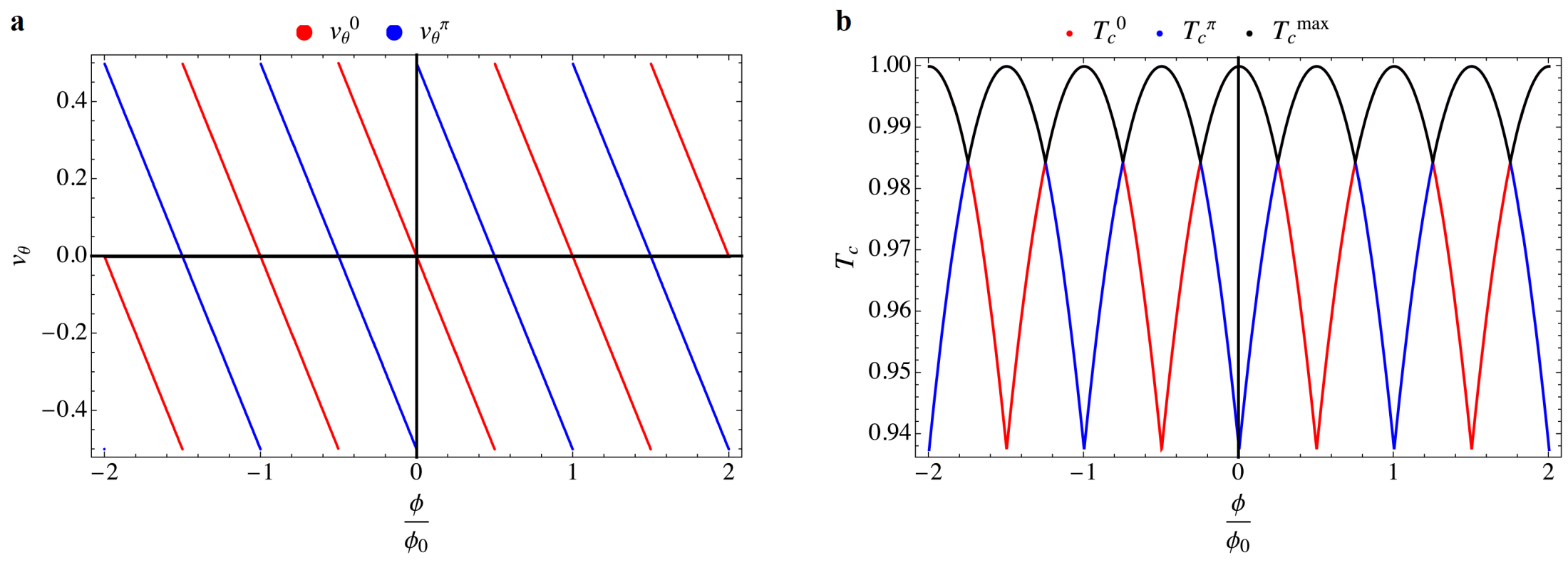

**Fig. S17**. Variation of the (a) circulating supercurrent velocity $v_\theta$, and (b) the superconducting order parameter $|\psi|^2$ as a function of the external magnetic flux $\frac{\Phi}{\Phi_0}$. The red and blue lines correspond respectively to the $\psi_0$ and $\psi_\pi$ *components*. Ginzburg-Landau parameters used for the two components are $T_{c_0}{}^{\pi} = T_{c_0}{}^{0} = 1$; $a_\pi = a_0 = 1$; $\beta_\pi = \beta_0 = 1$; $R = 1$.

By rewriting the free energy at fixed $v_\theta$,

$$f = \alpha|\psi|^2 + \frac{\beta}{2}|\psi|^4 + \frac{R^2}{4K}|\psi|^2 v_\theta^2,$$

we find the corresponding reduction of the critical temperature as a function of $v_\theta$

$$T_c = T_c^0 - \frac{R^2}{4Ka} v_\theta^2.$$

The variation as a function of the external magnetic flux of $v_\theta$ and $T_c$ are shown in Fig. S17 (a) and (b) both for the $\psi_0$ (red lines) and $\psi_\pi$ (blue lines) components. In Fig. S17(b), we also show the profile of the maximal critical temperature in a scenario where the two components coexist and have the same bare critical temperature. In this case, h/4e oscillations naturally emerge.

In addition to circulating current, we now introduce an external bias current and look at the fate of the Little-Parks effect.

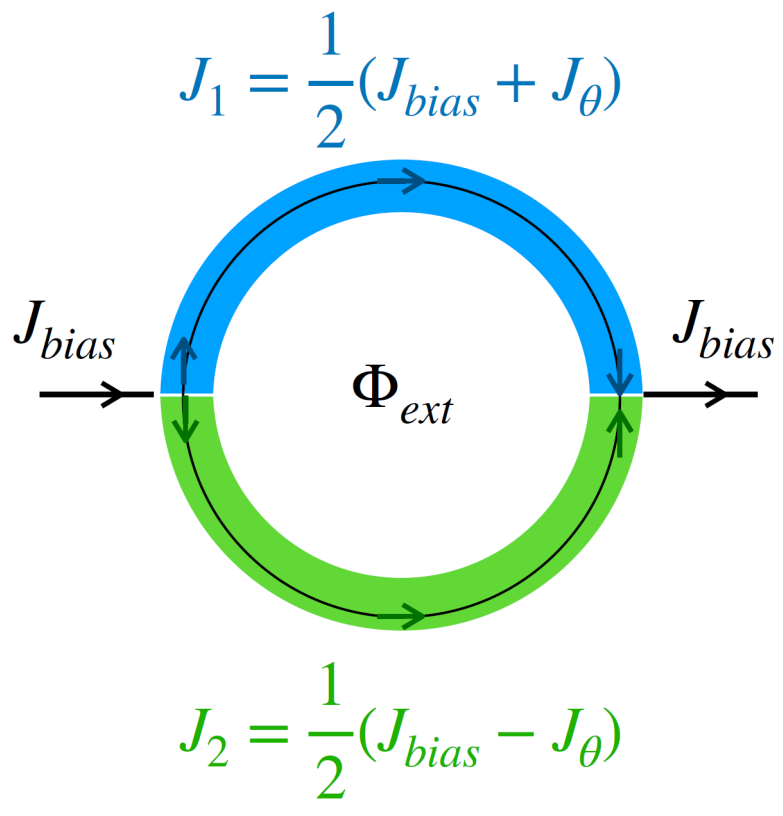

**Fig. S18. Schematic illustration of the ring**. Here we assume the current in the two arms to be different and equal to $J_1 = \frac{1}{2}(J_{bias} + J_\theta)$ and $J_2 = \frac{1}{2}(J_{bias} - J_\theta)$.

As schematically illustrated in Fig. S18, we label $J_1$ and $J_2$ the current density in the upper and lower arm respectively. The bias and the circulating current are defined as

$$J_\theta = J_1 - J_2;$$

$$J_{bias} = J_1 + J_2;$$

so that

$$J_1 = \frac{1}{2}(J_{bias} + J_\theta) = \frac{e|\psi|^2}{2}(v_s + v_\theta);$$

$$J_2 = \frac{1}{2}(J_{bias} - J_\theta) = \frac{e|\psi|^2}{2}(v_s - v_\theta).$$

The free energy of the ring can be written as the sum of the two contributions associated with the two arms of the ring as

$$f = \alpha|\psi|^2 + \frac{\beta}{2}|\psi|^4 + \frac{R^2}{8K}|\psi|^2(v_s + v_\theta)^2 + \frac{R^2}{8K}|\psi|^2(v_s - v_\theta)^2,$$

which in the case of two identical arms simplifies to

$$f = \alpha|\psi|^2 + \frac{\beta}{2}|\psi|^4 + \frac{R^2}{4K}|\psi|^2\left(v_s^2 + v_\theta^2\right).$$

By minimizing as before the free energy with respect to $|\psi|$ at fixed value of $v_s$ and $v_\theta$, we find

$$|\psi|^2 = |\psi(0)|^2\left[1 - \frac{R^2}{4|\alpha|K}\left({v_\theta}^2 + v_s^2\right)\right].$$

The bias current density then reads

$$J_{bias} = e|\psi(0)|^2\left[1 - \frac{R^2}{4|\alpha|K}\left({v_\theta}^2 + v_s^2\right)\right]v_s.$$

From the above expression, we identify the critical bias current of the ring as the maximum of the current as a function of $v_s$. In the presence of a finite magnetic flux, i.e. when $v_\theta \neq 0$, $J_{bias}$ is maximized by ${v_s}^2 = \frac{4|\alpha|K}{3R^2} - \frac{v_\theta^2}{3}$ and reads

$$J^c{}_{bias} = \frac{2e}{3}|\psi(0)|^2\left[1 - \frac{R^2{v_\theta}^2}{4|\alpha|K}\right]\left(\frac{4|\alpha|K}{3R^2} - \frac{v_\theta^2}{3}\right)^{1/2}.$$

This can be conveniently rewritten as

$$J^c{}_{bias}(T,\Phi) = J^c{}_{bias}(0,\Phi)\left(1-\frac{T}{T_c(0,\Phi)}\right)^{3/2},$$

where the critical current at zero temperature is $J^c{}_{bias}(0,\Phi) = \frac{2e}{3\,\beta}\left(\frac{4\mathrm{K}}{3R^2}\right)^{1/2} a^2 T_c(0,\Phi)^2$, and $T_c(0,\Phi)$ is the critical temperature at finite external magnetic flux and zero current. The above relation, can be easily inverted to find the temperature at which a given bias current, $J_{bias}$, becomes the critical current,

$$T_c(J_{bias},\Phi\,) = T_c(0,\Phi)\left(1-\frac{J_{bias}}{J^c{}_{bias}(0,\Phi)}\right)^{2/3}.$$

Both $T_c(\,0,\Phi)$ and $J_c(0,\Phi)$ depend on the specific superconducting component, e.g. via their coherence and London penetration lengths, so in general the dominant and the subdominant superconducting component will have different critical temperatures at zero current and different critical currents at zero temperature for a given external magnetic flux.

In Fig. S19 (a) and (b), we show for $\frac{\Phi}{\Phi_0} = 0$ and $\frac{\Phi}{\Phi_0} = 0.5$ respectively, the values of the two maximal critical temperatures for the two superconducting components as a function of the bias current density. For the two plots, one can see that it is possible to realize a scenario where at any applied magnetic field there exists a finite value of the bias current, $J^*(\Phi)$, at which the relative magnitude of the two critical temperatures is reversed. In Fig. S19 (c), we show the maximum critical temperature of the two-component model as a function of the external magnetic flux and for different values of the bias current, while Fig. S19 (d) presents a density plot of the maximum critical temperature as a function of $\frac{\Phi}{\Phi_0}$ and $J_{bias}$. One can see that by approaching $J^*(\Phi)$, the $h/2e$ $\pi -$shift oscillations coming from the dominant $\psi_\pi$ order parameter$-$ overlaps with the $h/2e$ oscillations of the subdominant $\psi_0$ order parameter, which becomes the dominant one at large enough bias current. The origin of the $h/4e$ oscillations is then a coexistence of the two solutions having comparable critical temperatures in a given current-bias range, as shown also in Fig. S17 (b).

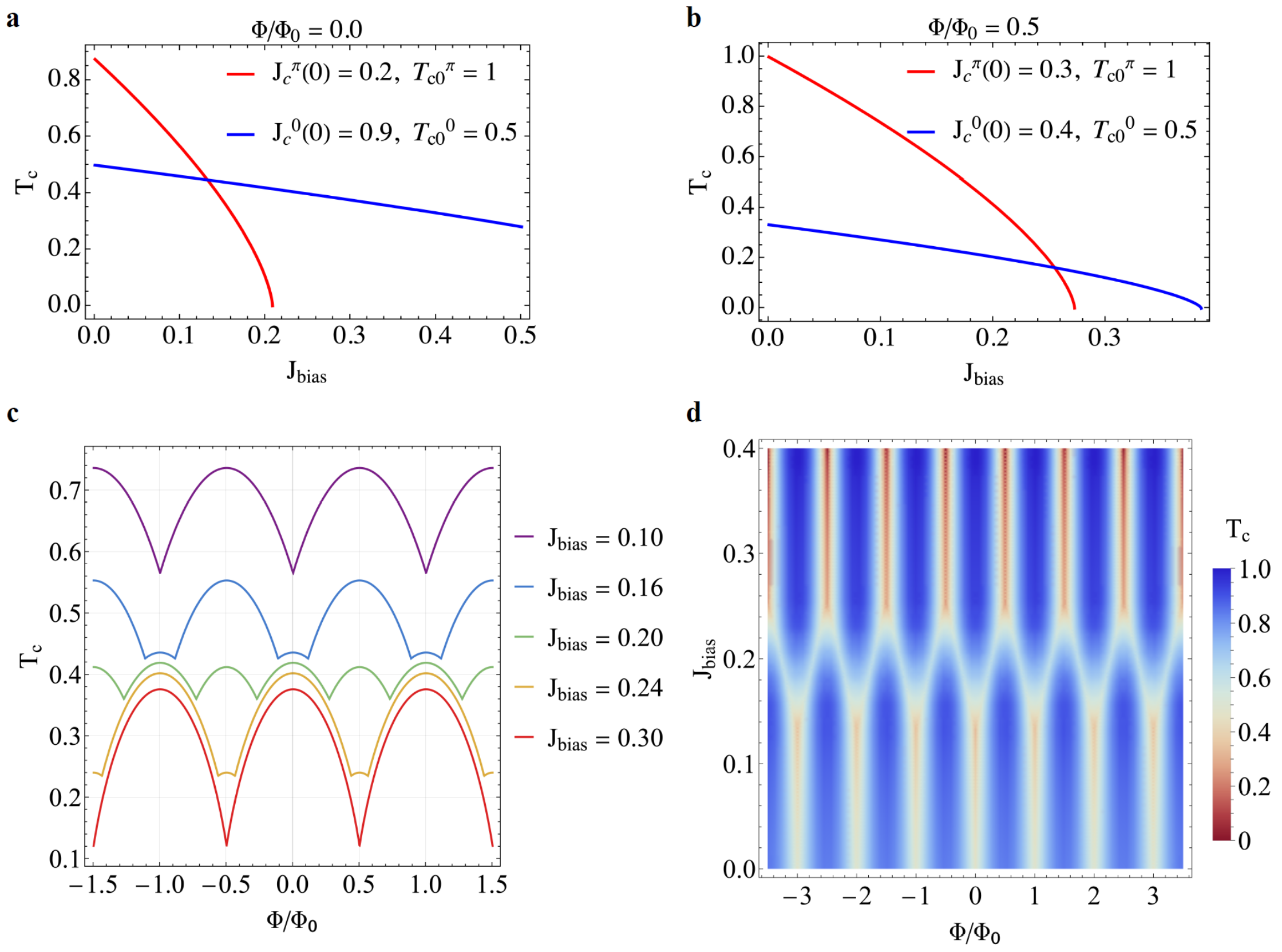

**Fig. S19 Theoretical results of the Little-Parks effect at different bias currents**. In panels (a) and (b), we show the critical temperature dependence of the two components as a function of the applied bias current. In panel (c), we report the maximum critical temperature of the two-component minimal model as a function of the external magnetic flux and for different values of the bias current. Finally, panel (d) shows the density plot of the maximum critical temperature as a function of $\frac{\Phi}{\Phi_0}$ and $J_{bias}$, where we subtracted the background decrease of the critical temperature with increasing current. For each value of $J_{\text{bias}}$, this background is defined as the average critical temperature over $\Phi/\Phi_0$. The Ginzburg-Landau parameters used for the two components are $T_{c_0}{}^{\pi} = 1$; $a_\pi = 1$; $\beta_\pi = 2$; $K_\pi = 0.5$; $T_{c_0}{}^{0} = 0.5$; $a_0 = 1.5$; $\beta_0 = 0.5$; $K_0 = 1; R = 1$.

## References

1. Wilson, S. D. & Ortiz, B. R. $AV_3Sb_5$ kagome superconductors. *Nat Rev Mater* **9**, 420–432 (2024).
2. Hossain, M. S. *et al.* Unconventional gapping behaviour in a kagome superconductor. *Nat. Phys.* **21**, 556–563 (2025).
3. Jang, J. *et al.* Observation of Half-Height Magnetization Steps in $Sr_2RuO_4$. *Science* **331**, 186–188 (2011).
4. Li, Y., Xu, X., Lee, M.-H., Chu, M.-W. & Chien, C. L. Observation of half-quantum flux in the unconventional superconductor β-$Bi_2Pd$. *Science* **366**, 238–241 (2019).
5. Xu, X., Li, Y. & Chien, C. L. Spin-Triplet Pairing State Evidenced by Half-Quantum Flux in a Noncentrosymmetric Superconductor. *Phys. Rev. Lett.* **124**, 167001 (2020).
6. Xu, X., Li, Y. & Chien, C. L. Observation of Odd-Parity Superconductivity with the Geshkenbein-Larkin-Barone Composite Rings. *Phys. Rev. Lett.* **132**, 056001 (2024).
7. Wan, Z. *et al.* Unconventional superconductivity in chiral molecule–$TaS_2$ hybrid superlattices. *Nature* **632**, 69–74 (2024).
8. Almoalem, A. *et al.* The observation of π-shifts in the Little-Parks effect in 4Hb-$TaS_2$. *Nat Commun* **15**, 4623 (2024).
9. Moll, P. J. W. Focused Ion Beam Microstructuring of Quantum Matter. *Annu. Rev. Condens. Matter Phys.* **9**, 147–162 (2018).
10. Höflich, K. *et al.* Roadmap for focused ion beam technologies. *Applied Physics Reviews* **10**, 041311 (2023).
11. Huff, M. Recent Advances in Reactive Ion Etching and Applications of High-Aspect-Ratio Microfabrication. *Micromachines* **12**, 991 (2021).
12. Guo, C. *et al.* Correlated order at the tipping point in the kagome metal $CsV_3Sb_5$. *Nat. Phys.* **20**, 579–584 (2024).
13. Little, W. A. & Parks, R. D. Observation of Quantum Periodicity in the Transition Temperature of a Superconducting Cylinder. *Phys. Rev. Lett.* **9**, 9–12 (1962).
14. Aoyama, K. Little-Parks oscillation and d-vector texture in spin-triplet superconducting rings with bias current. *Phys. Rev. B* **106**, L060502 (2022).
15. Aoyama, K. Half-quantum flux in spin-triplet superconducting rings with bias current. *Phys. Rev. B* **108**, L060502 (2023).
16. Fischer, M. H., Lee, P. A. & Ruhman, J. Mechanism for π phase shifts in Little-Parks experiments: Application to 4Hb−$TaS_2$ and to 2H−$TaS_2$ intercalated with chiral molecules. *Phys. Rev. B* **108**, L180505 (2023).
17. Michael Tinkham. *Introduction to Superconductivity*.